\documentclass[aps,prx,twocolumn, nofootinbib, epsfig,showpacs,floatfix, bibliography]{revtex4-2}
\usepackage{amsmath}
\usepackage{xcolor}
\usepackage{url}
\usepackage{physics}
\usepackage{graphicx}
\usepackage{amsfonts}
\usepackage{mathrsfs}
\usepackage{amsmath}
\usepackage{amssymb}
\usepackage{float}
\usepackage{bbold}
\usepackage[T1]{fontenc}
\usepackage[ttdefault=true]{AnonymousPro}
\usepackage[section]{placeins}
\usepackage[utf8x]{inputenc}
\usepackage{times}
\usepackage[colorlinks=true, linkcolor=blue, urlcolor=blue, citecolor=blue]{hyperref}
\makeatletter
\newcommand*{\rom}[1]{\expandafter\@slowromancap\romannumeral #1@}
\makeatother
\usepackage{multirow}
\hypersetup{
  colorlinks=true,
  citecolor=blue,
  linkcolor=blue,
  urlcolor=blue}

\begin{document} 

\title{ Work Statistics and Quantum Trajectories:\\ No-Click Limit and Non-Hermitian Hamiltonians }
 
\author{Manali Malakar, Alessandro Silva}

\affiliation{International School for Advanced Studies (SISSA), Via Bonomea 265, 34136 Trieste, Italy}

\date{\today}

\begin{abstract}
We investigate quantum work statistics within the standard two-point measurement (TPM) scheme in continuously monitored quantum systems, including the effects of generalized unitary evolution, possibly controlled by quantum circuit models, and  multiple generalized measurements as well as post-selection of \it no-click \rm trajectories. We derive an explicit expression for the work generating function that naturally incorporates non-Hermitian dynamics arising from quantum jump processes and reveals deviations from the standard Jarzynski equality due to measurement-induced asymmetries. We illustrate our theoretical framework by analyzing a one-dimensional transverse-field Ising model under local spin monitoring. In this model, increased measurement strength projects the system onto the \it no-click \rm state, leading to a suppression of energy fluctuations and measurement-induced energy saturation, reminiscent of the quantum Zeno effect. Moreover, we find signatures of the measurement induced transition observed in the no-click limit in the moments of the work distribution.
\end{abstract}

\maketitle

\section{Introduction}
\label{introduction}
Quantum thermodynamics is a rapidly developing interdisciplinary field, bridging quantum mechanics, statistical physics, and information theory, investigating how thermodynamic concepts extend to quantum systems operating far from equilibrium \cite{gemmer, anders, kosloff, talkner, silva}. A central aspect of quantum thermodynamics involves the extension of classical stochastic thermodynamics~\cite{Seifert_2012} to account for quantum fluctuations, i.e. fluctuations of entropy, heat, and work at the quantum level, where quantum coherence, entanglement, and measurement backaction become fundamentally relevant \cite{mukamel, jacobs, goold, adesso}. Both classical fluctuation theorems and constraints such as the Jarzynski equalities can be extended to the quantum regime, provided that the definitions of heat, entropy, and work are appropriately refined in the quantum context. For example, focusing on quantum work statistics, one typically employs the two-point measurement scheme, which relies on initial and final energy measurements after an Hamiltonian evolution \cite{kurchan, talkner, mukamel, tasaki2000}. 
Starting with the formulation above one can extend it to account for other generalized measurements during the evolution~\cite{Watanabe14,Watanabe2,PrasannaVenkatesh_2014}, to evolutions generated by unital and non-unital channels~\cite{Rastegin_2013,Rastegin14,Albash13,Aberg18}, as well as to special classes of non-Hermitian systems~\cite{Deffner15,Deffner16,Zeng_2017,Wei18}.\\~\\ 
Focusing on non-Hermitian systems previous research on the subject was devoted to the classification of the conditions under which the Jarzynski equalities~\cite{Deffner15,Wei18}, thermodynamic bounds~\cite{Deffner16} and fluctuation theorems~\cite{Wei18,Zeng_2017} are still valid. 
Pseudo-Hermiticity and $\mathcal{PT}$-symmetry~\cite{ueda} appear to be the key to maintain validity of quantum thermodynamics in the standard formulation. It is however important to notice that frequently non-Hermitian Hamiltonians describe the evolution of open systems subject to continuous measurement~\cite{milburn} once we post-select a special, \it no-click \rm trajectory~\cite{turkeshi_5,turkeshi_4}. In this setting, instead of being guided by the consistency of the defintion with fluctuation theorems, one may start directly from the definition of quantum trajectories~\cite{Elouard17} and the two-point measurement scheme to construct a notion of work statistics for non-Hermitian systems from scratch.\\~\\ 
The purpose of this work is to investigate quantum work statistics (or more appropriately two-point energy statistics) for continuously monitored quantum systems, focusing in particular on \it no-click \rm trajectories and their naturally arising non-Hermitian Hamiltonians. Our goal is to derive compact expressions for the  work generating function in the non-Hermitian case arising from quantum jump processes and subsequent post-selection. Introducing trajectories in a way similar to Ref.~[\onlinecite{Elouard17}], we first clarify how to formulate appropriately work statistics in the presence of quantum trajectories generated by a generic unitary dynamics (including quantum circuits~\cite{Skinner19}) and interspersed generalized measurement.  Once a \it no-click \rm trajectory is singled out, we discuss what are the expected modifications on the Jarzynski equality~\cite{Jarzynski97}. Finally, we demonstrate our formalism by analyzing the work statistics for a one-dimensional transverse-field quantum Ising model subject to local spin monitoring, a model that was shown to display a measurement induced phase transition (MIPT) also in the \textit{no-click} limit \cite{turkeshi_5}. On one hand our study reveals  phenomena such as measurement-induced energy saturation and reduction in fluctuations, reflecting the effective confinement of the system to specific quantum states. These effects, reminiscent of the quantum Zeno effect \cite{mishra}, highlight how continuous observation can effectively freeze the system dynamics, drastically altering the energy exchange processes and associated fluctuations. On the other hand, we observe signature of the MIPT occurring in the steady state entanglement as singularities of the average work done at the pohase boundary, revealing the potential of this quantity to reveal interesting many-body effects.\\~\\
The rest of the paper is organized as follows. In Sec.~\ref{theory}, we introduce the notion of quantum work statistics for quantum trajectories, subsequently focusing on quantum jumps and the \it no-click \rm trajectory described by non-Hermitian quantum dynamics. In Sec.~\ref{ising result}, we present our analysis and results for the quantum work statistics in a monitored transverse-field Ising chain. Finally, we summarize our findings and conclude in Sec.~\ref{conclusion}.

\section{Work statistics, trajectories and the no-click limit}
\label{theory}
Our first goal is to formulate TPM work statistics~\cite{Lutz07,Campisi11} for monitored dynamics with intermediate generalized measurements, and then to specialize the resulting expressions to the post-selected \textit{no-click} trajectory, for which the conditioned evolution is generated by an effective non-Hermitian Hamiltonian~\cite{ueda0}. In principle, for systems subject to Hamiltonian dynamics and generalized measurements, the quantity we compute should be called two-point energy statistics, and particular attention should be paid to defining work and heat appropriately along the trajectories~\cite{Elouard17}. Below we will be dealing, however, with generalized unitary dynamics as well, possibly generated by a unitary circuit of gates acting on a qubit register; therefore in order to make contact with previous literature on the subject, in particular on non-Hermitian problems~\cite{Deffner16,Deffner15,Wei18,Zhou21}, we will use the name "work statistics" throughout the paper.\\~\\
While previous attempts to define such quantity focused either on very specific non-Hermitian Hamiltonians ($\mathcal{PT}$ symmetric with real spectrum)~\cite{Deffner16,Deffner15,Wei18} or on the modification of the standard two measurement scheme~\cite{Zhou21}, the path that we will follow will be different. Indeed, one way to realize systematically the evolution with a non-Hermitian Hamiltonian is to consider a system subject to quantum jumps and postselect the so-called \it no-click \rm trajectory~\cite{milburn,turkeshi_5}, the exponentially rare trajectory that corresponds to a null result  at every measurement attempt. Defining work statistics for such a trajectory requires extending the notion of work statistics to systems subject to multiple generalized measurements~\cite{Watanabe14,Gherardini22}. Once this is done the work statistics for a non-Hermitian evolution is obtained by just post selecting the contribution of the \it no-click \rm trajectory. This is the overall direction that we will follow in this section, where we also discuss in detail step by step the validity and significance of the constraint given by the Jarzynski equality~\cite{Jarzynski97,Campisi11} and why we should not expect, in general, the work statistics associated to a non-Hermitian evolution to satisfy such constraint. 

\subsection{Work statistics and generalized measurements}~\label{sec1.1}

The standard protocol to define work statistics in thermally isolated quantum systems consists of taking the system along a quantum trajectory defined by an initial and a final energy measurement (relative to initial $H_i$ and final $H_f$ Hamiltonians). If $p_i(n)$ and $p_f(m|n)$ are the probability to measure the initial energy $E_i(n)$ and final one $E_f(m)$ (conditional to the initial energy measurement) the work probability distribution is defined as \cite{kurchan, Jarzynski97, Lutz07}:
\begin{eqnarray}~\label{ws}
P(W)=\sum_{n,m}p_i(n)p_f(m|n)\delta\big(W\!-\![E_f(m)\!-\!E_i(n)]\big)   
\end{eqnarray}
The initial measurement is usually taken with respect to an equilibrium distribution at inverse temperature $\beta$, i.e. $p_i(n)=\exp[-\beta E_i(n)]/Z_i$, where $Z_i$ is the initial partition function. The conditional probability $p_f(m|n)$ in turn depends on the specifics of the evolution and of the processes occurring between the two measurements.\\~\\
For a generic coherent evolution represented by the unitary operator 
$U_{t_f,t_i}$, we have ~\cite{Lutz07}
\begin{eqnarray}
  p_f(m|n)=|\langle \psi_f(m)| U_{t_f,t_i}|\psi_i(n)\rangle|^2.  
\end{eqnarray}
Coherent dynamics can be generated through either a unitary quantum circuit~\cite{Skinner19} or standard Hamiltonian dynamics. In the latter case, if a parameter $\lambda(t)$ is varied in time between an initial $\lambda_i=\lambda(t_i)$ and a final $\lambda_f=\lambda(t_f)$ value, we have
\begin{eqnarray}\label{quench}
 U_{t,t_0}=\mathcal{T}{\rm exp}\bigg(-i\int_{t_0}^{t} dt'H(\lambda(t'))\bigg).   
\end{eqnarray}
where $\mathcal{T}$ represents the time-ordering operator. 
In the case of unitary circuits $U$
 will be defined by a sequence of unitary gates acting on a register of qbits.
If during the evolution the system is also subject to a single measurement of some quantity with possible outcomes $\{r\}$, the expression above changes~\cite{Watanabe14}. Let us first represent the \it generalized \rm measurement by a set of measurement 
operators $M_r$ subject to the constraint $\sum_r M^{\dagger}_rM_r=\mathbb{1}$. The conditional probability $p_f(m|r|n)$ to obtain $E_f(m)$ as the final energy measurement, given that $E_i(n)$  was obtained in the initial measurement and $r$ in the intermediate one, is:
\begin{eqnarray}
p_f(m|r|n)=\frac{|\langle \psi_f(m)|U_{t_f,t_r}M_rU_{t_r,t_i}|\psi_i(n)\rangle|^2 }{p(r|n)},   
\end{eqnarray}
where $t_r$ is the time at which the mid measurement occurs, and 
\begin{eqnarray}
p(r|n)=\langle \psi_i(n) |U^{\dagger}_{t_r,t_i}M^{\dagger}_r M_rU_{t_r,t_i}| \psi_i(n)\rangle,     
\end{eqnarray}\\
is the probability of obtaining outcome $r$ given $E_i(n)$ as the initial energy measurement. Since work statistics does not keep track of the result of the intermediate measurement the quantity entering in Eq.~(\ref{ws}) is the unconditional probability distribution
\begin{eqnarray}~\label{conditionalwithmeas}
 p_f(m|n)&=&\sum_r p(r|n)p_f(m|r|n)=\nonumber \\
 && \sum_r |\langle \psi_f(m)|T_{t_f,t_i}(r,t_r)|\psi_i(n)\rangle|^2,
\end{eqnarray}
where we introduced the operator $T_{t_f,t_i}(r,t_r)=U_{t_f,t_r}M_rU_{t_r,t_i}$.\\

Generalizing the expressions above to the case where instead of a single measurement we have multiple measurements with possible outcomes $\{ r_j\}$ at times $\{t_j\}$ ($j=1,\dots,N$), Eq.~(\ref{conditionalwithmeas}) becomes
\begin{eqnarray}~\label{conditionalgeneral}
 p_f(m|n)= \sum_{\{r_j\}} |\langle \psi_f(m)|T_{t_f,t_i}(\{r_j,t_j\})|\psi_i(n)\rangle|^2,
\end{eqnarray}
where now
\begin{eqnarray}~\label{generalT}
 T_{t_f,t_i}(\{r_i,t_i\})=U_{t_f,t_N}M_{r_N}U_{t_N,t_{N-1}}
 \dots M_{r_1}U_{t_1,t_i}.
\end{eqnarray}
The combination of Eq.~(\ref{ws}) with the expression for the conditional probability in Eq.~(\ref{conditionalgeneral}), in terms of the operator $T_{t_f,t_i}$ given by  Eq.~(\ref{generalT}), constitutes the most general expression for work statistics for quantum trajectories generated by generalized measurements and generic unitary dynamics. 

\subsection{Generating function and Jarzynski equalities}~\label{Jarzynski}

In order to further extend our analysis and specialize it to the case of \it no-click \rm trajectories, let us now consider the generating function of work statistics, given by:
\begin{eqnarray}
 {\cal G}(u)=\int dW P(W) e^{-iWu}.   
\end{eqnarray}
Using Eq.(\ref{ws})-(\ref{conditionalgeneral})-(\ref{generalT}) one obtains
\begin{widetext}
\begin{eqnarray}\label{genfun}
{\cal G}(u)=\sum_{\{r_j\}} {\rm Tr}\left[ T^\dagger_{t_f,t_i}(\{r_j,t_j\})\;e^{-iH_f u}\; T_{t_f,t_i}(\{r_j,t_j\})\; e^{iH_iu}\frac{e^{-\beta H_i}}{Z_i}\right], 
\end{eqnarray}
\end{widetext}
which is the  general expression for the generating function of work statistics. While the identity 

\begin{eqnarray}\label{norma1}
 \sum_{\{r_j\}}  T^\dagger_{t_f,t_i}(\{r_j,t_j\})T_{t_f,t_i}(\{r_j,t_j\}) 
 ={\mathbb 1},
\end{eqnarray}
guarantees the normalization of the probability distribution $P(W)$, since
${\cal G}(u=0)=\int dW P(W)=1$, the Jarzynski equality is obtained 
by setting
\begin{eqnarray}
{\cal G}(-i\beta)=\frac{{\rm Tr}\left[ \left(\sum_{\{r_j\}} T_{t_f,t_i}(\{r_j,t_j\})T^\dagger_{t_f,t_i}(\{r_j,t_j\}) \right) e^{-\beta H_f} \right]}{Z_i}\nonumber
\end{eqnarray}
from which we see that, provided~\cite{Rastegin_2013}

\begin{eqnarray}\label{norma2}
 \sum_{\{r_j\}}  T_{t_f,t_i}(\{r,t_i\})T^{\dagger}_{t_f,t_i}(\{r,t_i\}) 
 ={\mathbb 1},
\end{eqnarray}

one obtains
\begin{eqnarray}
    \langle e^{-\beta W} \rangle = {\cal G}(-i\beta)= \frac{Z_f}{Z_i}=e^{-\beta \Delta F}.
\end{eqnarray}
Eq.~(\ref{norma2}) is the condition of unitality of the quantum channel, representing the dynamics of our system~\cite{Rastegin_2013,Rastegin14,Albash13}. 
Let us now discuss the physical meaning of the condition. Unitality is obvious for unitary evolutions while the presence of generalized measurements implies that the constraint in Eq.~(\ref{norma2}) 
is a consequence of the requirement that measurement operators satisfy
\begin{eqnarray}\label{constraint}
 \sum_r M_rM^{\dagger}_r={\mathbb 1}.   
\end{eqnarray}
Since a measurement with outcome $r$ modifies the density matrix
$\rho_0$ according to the relation
\begin{eqnarray}
 \rho_r=\frac{M_r\rho_0 M^\dagger_r}{p_r},   
\end{eqnarray}
where $p_r=\langle M^\dagger_rM_r \rangle$, the condition in Eq.~(\ref{norma2}) implies that the mapping for \it unconditional \rm evolution of a density matrix after a single measurement
\begin{eqnarray}
\rho=\sum_r p_r \rho_r=   \sum_r M_r\rho_0 M^\dagger_r,
\end{eqnarray}
is itself \it unital\rm.\\~\\
While most of the measurement operators (projectors, quantum diffusion, quantum jumps, etc.) do satisfy these conditions, it does not come as a surprise that one can easily find a set that does not.
Consider, for example, a set of measurement outcomes $r=0,\dots,M$, and
associate to them the states $| r\rangle$ in the Hilbert space ($\sum_{r}|r \rangle \langle r |=\mathbb{1}$). The operators 

\begin{eqnarray}
  M_r=|0\rangle \langle r| , 
\end{eqnarray}
describing a measurement after which the state is reset to $|0\rangle$ regardless of the outcome, satisfy the normalization condition of measurement operators but not Eq.~(\ref{constraint}).

\subsection{No-click trajectories and non-Hermitian physics}\label{nonH}

We are now in the position to specialize Eq.~(\ref{genfun}) to the case of quantum jumps and, in particular, to \it no-click \rm trajectories. Following Ref.~\cite{wiseman}, let us discretize the time span $[0,t]$ into $N$ small steps of duration $\delta t$, such that $\delta t=t/N$ and perform, in each interval, a measurement with the generalized operators
\begin{eqnarray}
 M_0(\delta t)&=&{\mathbb 1}-\left(\frac{R}{2}+iH\right)\delta t,\label{kraus_noclick}\\
  M_1(\delta t)&=&\sqrt{\delta t}\;c,
\end{eqnarray}
where $R=c^\dagger c$ and $H$ is an Hermitian operator, representing the unitary evolution. Notice that the condition
in Eq.~(\ref{constraint}) is satisfied provided $c^{\dagger} c=cc^{\dagger}$, which is the case, for example, when $c=|1\rangle \langle 1 |$ is a projector.\\~\\
It is now rather straightforward to write down the generating function specialized to this set of operators. Our goal now is to specialize the work statistics to a single trajectory; that is, rather than computing the unconditional $P(W)$, we focus on the trajectory where the measurement gives systematically no result.
In this case, we specialize Eq.~(\ref{generalT}) to the \textit{no-click} record, i.e., $r_j=0$ for all $j=1, \dots, N$, and assume equally spaced monitoring times $t_j$. As the \textit{no-click} Kraus operator $M_{0}(\delta t)$ in Eq.~(\ref{kraus_noclick}) already incorporates the system's Hermitian Hamiltonian $H$, the unitary segments $U_{t_j, t_{j+1}}=e^{-i H \delta t}$ appearing in Eq.~(\ref{generalT}) are effectively absorbed into $M_{0}(\delta t)$~\cite{milburn,wiseman}. With this convention and taking the continuum limit $N \rightarrow \infty$ ($\delta t \rightarrow 0$), the evolution operator for the \textit{no-click} trajectory is given by:
\begin{eqnarray}\label{Tnoclick}
T_{\{0\}}&=& \underset{N \rightarrow \infty}{\rm lim} \big[M_0(\delta t)\big]^{N}\notag\\
&=& \underset{N \rightarrow \infty}{\rm lim} \bigg[\mathbb{1}-\bigg(\frac{R}{2}+ i H\bigg)\frac{t}{N}\bigg]^{N}\notag\\
&=& e^{-i H_{\rm eff}t}
\end{eqnarray}
where $H_{\rm eff}=(H-i R/2)$. The last equality follows from the standard limit definition of the matrix exponential.\\~\\
Therefore the generating function of work statistics for the \it no-click \rm trajectory is:
\begin{eqnarray}\label{genfunc_0}
{\cal G}_{\{0\}}=\frac{{\rm Tr} \left[ e^{i H^\dagger_{{\rm eff}}t} e^{-iH_f u}e^{-i H_{{\rm eff}} t} e^{i H_iu}\rho_i \right]}{ {\rm Tr} \left[ e^{i H^\dagger_{{\rm eff}}t} e^{-i H_{{\rm eff}} t}\rho_i \right]},   
\end{eqnarray}
where we normalize by the probability of occurrence of the \it no-click \rm trajectory. While we still have ${\cal G}_{\{0\}}(u=0)=1$, the Jarzynski equality is not satisfied; instead, we have:

\begin{eqnarray}
 {\cal G}_{\{0\}}(-i\beta)= \frac{{\rm Tr} \left[ e^{-i H_{{\rm eff}} t} e^{i H^\dagger_{{\rm eff}}t} e^{-\beta H_f}\right]}{ {\rm Tr} \left[ e^{i H^\dagger_{{\rm eff}}t} e^{-i H_{{\rm eff}} t}e^{-\beta H_i} \right]},
\end{eqnarray}
which is formally similar but does not correspond to a thermodynamic equilibrium quantity. We can nevertheless cast it in a familiar form
\begin{eqnarray}
\langle e^{-\beta (W-\Delta F)}\rangle = \gamma_t, 
\end{eqnarray}
where the \it efficacy \rm~\cite{Albash13} is:
\begin{eqnarray}\label{efficacy}
   \gamma_t=\frac{\langle e^{-i H_{{\rm eff}} t} e^{i H^\dagger_{{\rm eff}}t}\rangle_f}   {\langle e^{i H^\dagger_{{\rm eff}} t} e^{-i H_{{\rm eff}}t}\rangle_i}, 
\end{eqnarray}
with $\langle \cdot \rangle_{f,i}={\rm Tr}[\;\cdot\;\rho_{f,i}]$. Notice that the efficacy characterizes the asymmetry between forward and backward evolution, arising from the non-commutativity of the Hamiltonian and its adjoint, i.e., $[H_{\rm eff},H^{\dagger}_{\rm eff}]\neq 0$.\\~\\
Although projective energy measurements are impractical, even in many-body systems the modified Jarzynski equality can still be tested by measuring the work generating function with a Ramsey interferometer using an ancilla qubit, then Fourier transforming it to obtain the work distribution. This approach was proposed in single-qubit interferometry schemes~\cite{Dorner2013,Mazzola2013} and demonstrated in NMR and trapped-ion experiments~\cite{Batalhao2014, An2015}. In our monitored setting the same interferometric protocol is run conditioned on a \textit{no-click} record of the continuous measurement. Concretely, we take an ancilla qubit $\mathcal{A}$ prepared in $(|0\rangle+|1\rangle)/\sqrt{2}$, and the system in $|\Psi_{0}\rangle$, the ground state of $H_{i}$. Postselecting the \textit{no-click} trajectories, the system evolves under the non-Hermitian propagator $e^{-iH_{\rm eff}t}$ for time $t$, so the joined state becomes
\begin{equation*}
|\Phi\rangle = \frac{|0\rangle \otimes e^{-iH_{\rm eff}t} |\Psi_{0}\rangle+|1\rangle \otimes e^{-iH_{\rm eff}t} |\Psi_{0}\rangle}{\sqrt{2\langle \Psi_0|e^{iH^{\dagger}_{\rm eff}t}  e^{-iH_{\rm eff}t}|\Psi_{0}\rangle}},
\end{equation*}
where the \textit{no-click} success probability $\langle \Psi_0|e^{iH^{\dagger}_{\rm eff}t}  e^{-iH_{\rm eff}t}|\Psi_{0}\rangle$ in the denominator normalizes $|\Phi\rangle$. Now applying a controlled operation $e^{-iH_{f}u}$ on the system, conditioned on ancilla $\mathcal{A}=|1\rangle$ gives
\begin{equation*}
|\Phi_{1}\rangle = \frac{|0\rangle \otimes e^{-iH_{\rm eff}t} |\Psi_{0}\rangle + |1\rangle \otimes e^{-iH_{f}u} e^{-iH_{\rm eff}t} |\Psi_{0}\rangle}{\sqrt{2\langle \Psi_0|e^{iH^{\dagger}_{\rm eff}t}  e^{-iH_{\rm eff}t}|\Psi_{0}\rangle}}.
\end{equation*}
A Ramsey readout on $\mathcal{A}$ then yields the postselected generating function
\begin{equation}
\mathcal{G}_{\{0\}}(u,t)= \frac{\langle \Psi_0|e^{iH_{\rm eff}^{\dagger}t} e^{-iH_{f}u} e^{-iH_{\rm eff}t}|\Psi_{0}\rangle}{\langle \Psi_0|e^{iH^{\dagger}_{\rm eff}t}  e^{-iH_{\rm eff}t}|\Psi_{0}\rangle} e^{iE^{i}_{0}u}.   
\end{equation}
The overall phase $e^{iE_0u}$ reflects the known action of $e^{iH_iu}$ on the prepared eigenstate and can be incorporated by post-processing.

\section{Results for monitored quantum Ising model}
\label{ising result}

We are now ready to study the work statistics for a one-dimensional transverse-field Ising model subject locally to the monitoring of the up component of the transverse spin. In the \it no-click \rm limit, the effective Hamiltonian describing the dynamics is: 
\begin{eqnarray}
H_{\rm eff}[h,\gamma]=-J\sum\limits_{i=1}^{L}\hat{\sigma}^{z}_{i}\hat{\sigma}^{z}_{i+1}-\left(h+i\frac{\gamma}{4}\right)\sum\limits_{i=1}^{L}\hat{\sigma}^{x}_{i},
\label{H_NC}
\end{eqnarray}
where $\hat{\sigma}^{\alpha}$ with $\alpha \in \{x, y, z\}$ denote the Pauli spin matrices, $h$ represents the transverse field, and $\gamma$ is the measurement rate. Using the Jordan-Wigner transformation, Eq.~\eqref{H_NC} can be diagonalized exactly in terms of free fermions~\cite{turkeshi_5}.
The model is known to display a MIPT at a critical value of $\gamma_c=4\sqrt{1-h^2}$ between a logarithmic entanglement phase ($\gamma<\gamma_c$) and an area law phase ($\gamma>\gamma_c$). Below we show that remonants of this criticality are shown also in work statistics.
\\~\\
In order to define work statistics, we must specify the \it initial \rm and \it final \rm Hamiltonians. In the following, we will, for simplicity, take  $H_i=H_f=H_{\rm eff}[h,\gamma=0]$, and therefore study the work statistics originating solely from the introduction of the measurement. If the dynamics starts from the ground state $|\Psi_0 \rangle$ of $H_i$, 
the characteristic function can be written as:
\begin{eqnarray}
\mathcal{G}_{\{0\}}(u,t)= \frac{\langle \Psi_0|e^{iH_{\rm eff}^{\dagger}t} e^{-iH_{f}u} e^{-iH_{\rm eff}t}e^{iH_{i}u}|\Psi_{0}\rangle}{\langle \Psi_0|e^{iH^{\dagger}_{\rm eff}t}  e^{-iH_{\rm eff}t}|\Psi_{0}\rangle }.
\label{G-function}
\end{eqnarray}
 Within the free fermionic framework, $H_{i}$ can be expressed in terms of the Bogoliubov quasiparticle operator $\hat{\eta}_{k}$ which diagonalizes it in $k$-space, as
\begin{eqnarray}
H_{i}= \sum_{k>0}\epsilon^{i}_{k}\big(\hat{\eta}^{\dagger}_{k} \hat{\eta}_{k}+\hat{\eta}^{\dagger}_{-k} \hat{\eta}_{-k}\big) +E^{i}_{0},
\label{H_i}    
\end{eqnarray}
where $E_0^i=-\sum_{k>0}\epsilon^{i}_{k}$ is the ground state energy and $\epsilon^{i}_{k}=2\sqrt{(h-J\cos{k})^{2}+J^{2}\sin^{2}{k}}$ denotes the dispersion of quasiparticles, while the vacuum $|\Psi_{0}\rangle$ satisfies the relation $\hat{\eta}_{k}|\Psi_{0}\rangle=0$.\\

We can proceed in a similar way both for $H_{\rm eff}$ and $H^{\dagger}_{\rm eff}$, which can be written as~\cite{turkeshi_5}:
\begin{subequations}
\begin{align}
H_{\rm eff} &= \sum_{k>0}\epsilon^{\rm eff}_k\big(\hat{\gamma}^{*}_{k} \hat{\gamma}_{k}+\hat{\gamma}^{*}_{-k} \hat{\gamma}_{-k}\big) +E^{\rm eff}_{0}, \label{H_nc_a}\\
H^{\dagger}_{\rm eff} &= \sum_{k>0}\epsilon^{\rm eff^*}_k\big(\hat{\widetilde{\gamma}}^{*}_{k} \hat{\widetilde{\gamma}}_{k}+\hat{\widetilde{\gamma}}^{*}_{-k} \hat{\widetilde{\gamma}}_{-k}\big) +(E^{\rm eff}_{0})^*.
\label{H_nc_b}
\end{align}
\end{subequations}
%%%%%%%%%%%%%%%%Figure 1%%%%%%%%%%%%%%%%%%%%%%
\begin{figure}[h]
\centering
\includegraphics[width=0.8\columnwidth]{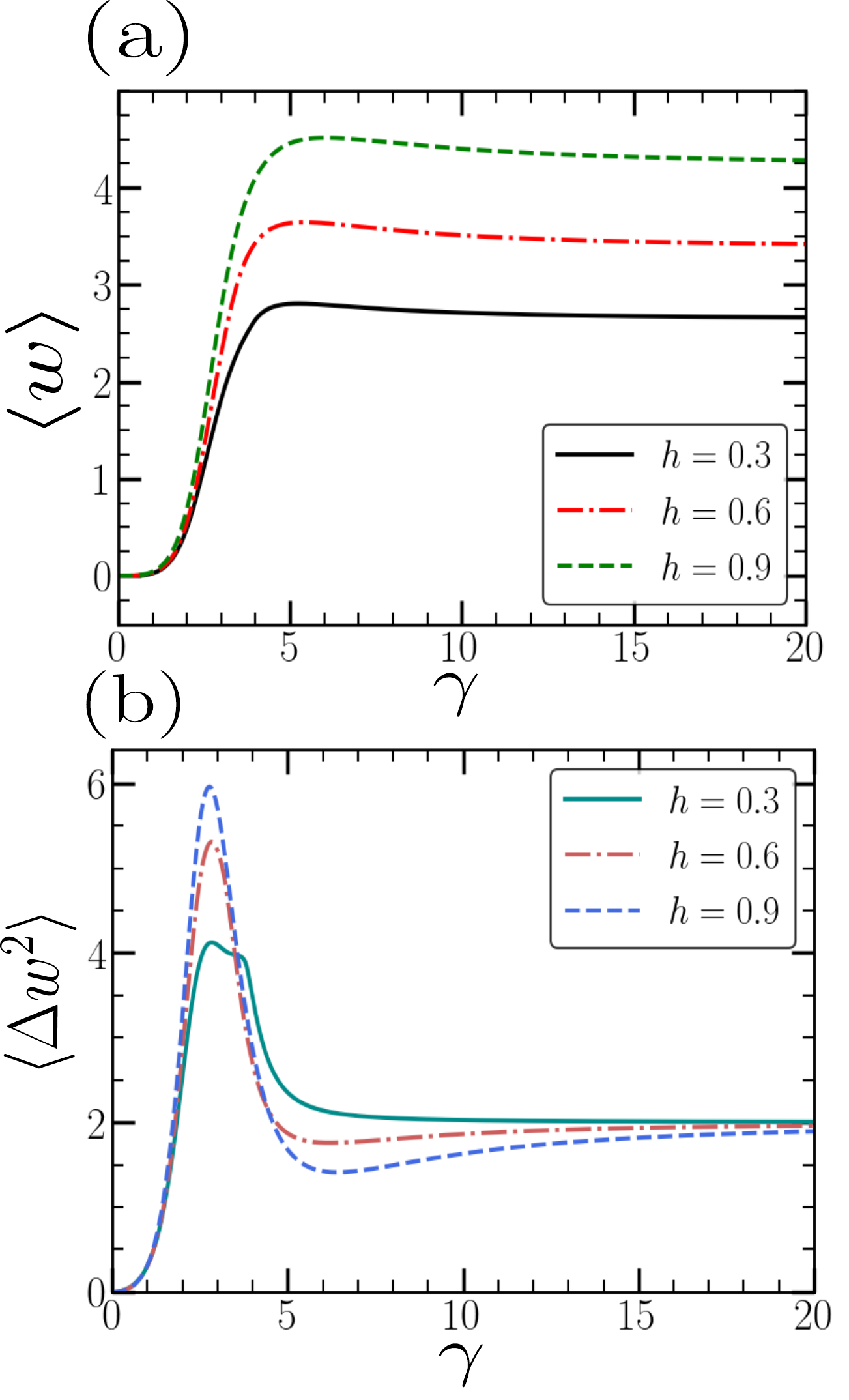}
\caption{(a) Average work density $\langle w \rangle$ (normalized by system size $L$) and (b) its variance $\langle \Delta w^2 \rangle$ plotted as functions of the measurement strength $\gamma$ for the monitored one-dimensional transverse-field Ising model at different values of the transverse field $h$ at time $t=1.0$. The plots demonstrate that when $\gamma=0$, both quantities are zero, reflecting the coincidence of the initial and effective non-Hermitian Hamiltonians. As $\gamma$ increases, the non-Hermitian dynamics progressively project the system onto the \textit{no-click} (spin-down) state, leading to a saturation of the work performed and a reduction in its fluctuations.}
\label{Fig1}
\end{figure}
%%%%%%%%%%%%%%%%%%%%%%%%%%%%%%%%%%%%%%%%%%%%%%
%%%%%%%%%%%%%%%%Figure 2%%%%%%%%%%%%%%%%%%%%%%
\begin{figure}[h!]
\centering
\includegraphics[width=0.85\columnwidth]{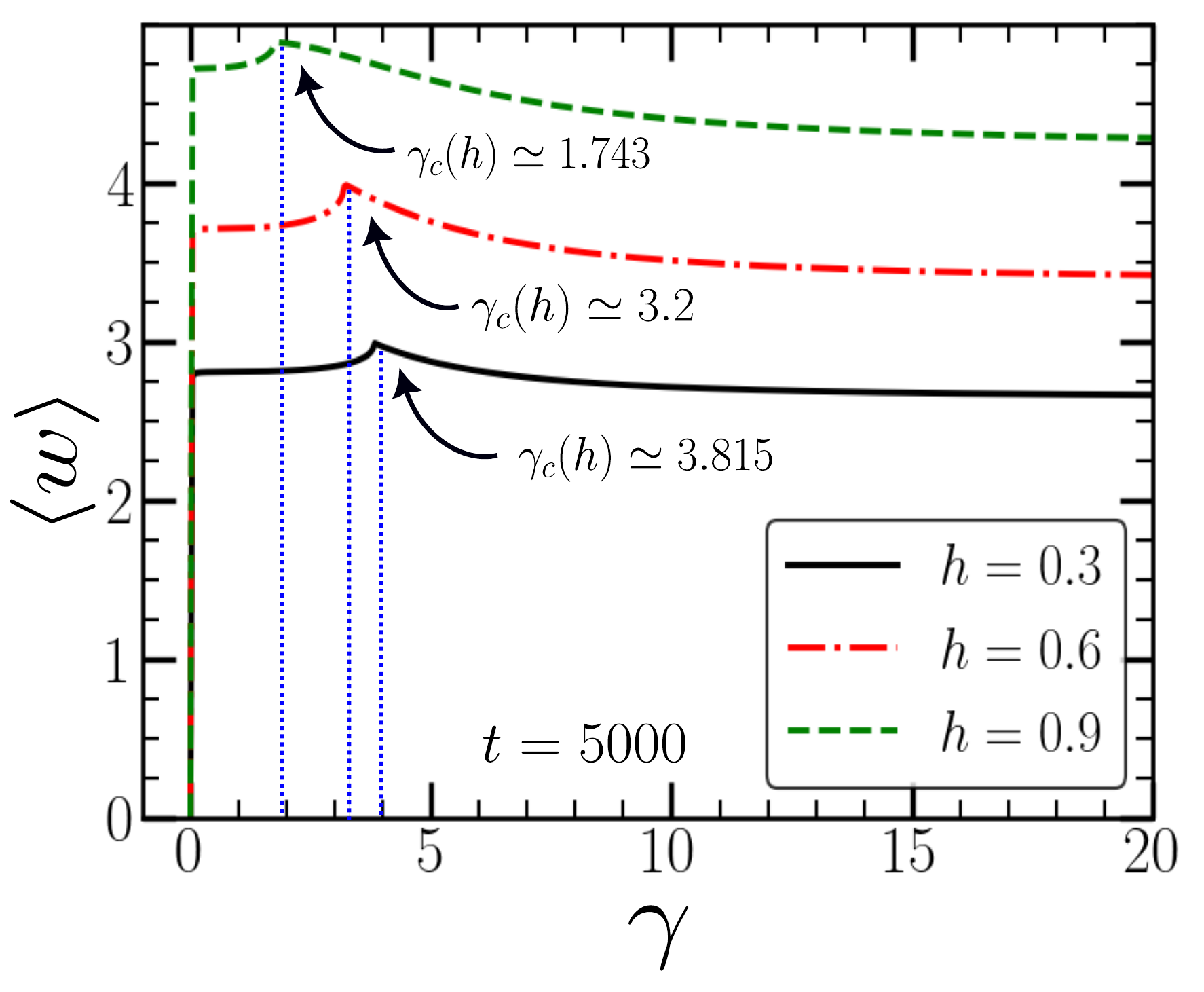}
\caption{Average work density $\langle w \rangle$ as a function of the measurement rate $\gamma$ for $h=\{0.3,0.6,0.9\}$ at long evolution time $t=5000$. A sharp rise appears at $\gamma \rightarrow 0^{+}$ when monitoring is first switched on. Each curve then shows a kink at a value of $\gamma$ that coincides with the analytic \textit{no-click} criticality $\gamma_{c}(h)=4\sqrt{1-h^2}$ (blue dotted guides and labels). For larger $\gamma$, $\langle w \rangle$ saturates due to Zeno suppression.}
\label{Fig2}
\end{figure}
%%%%%%%%%%%%%%%%%%%%%%%%%%%%%%%%%%%%%%%%%%%%%%
The complex eigenvalues are given by $\epsilon^{\rm eff}_k=2\sqrt{(h-J\cos{k}+i \gamma/4)^2+J^2\sin^{2}{k}}\equiv \lambda_{k}+i \Gamma_{k}$ with non-positive imaginary part $\Gamma_k \leq 0$, $\epsilon^{\rm eff^*}_k$ being its complex conjugate. Here, $\hat{\gamma}_{k}$ and $\hat{\widetilde{\gamma}}_{k}$ represent the non-Hermitian quasiparticles satisfying, $\hat{\gamma}_{k}|\emptyset \rangle=0$ and $\langle \widetilde{\emptyset}|\hat{\widetilde{\gamma}}^{*}_{k}=0$, where $|\emptyset \rangle$ and $\langle \widetilde{\emptyset}|$ denote the right and left non-Hermitian vacuum states, respectively.\\~\\
Our first aim is to express $|\Psi_{0}\rangle$ in terms of the non-Hermitian quasiparticle $\hat{\gamma}_{k}$ that diagonalizes $H_{\rm eff}$. This can be achieved by rewriting $\hat{\eta}_{k}$ in terms of $\hat{\gamma}_{k}$ (see Appendix.~\ref{AppendixA} for details) as follows:
\begin{eqnarray}
\hat{\eta}_{k}=X_{k}\hat{\gamma}_{k}-i Y_{k}\hat{\gamma}^{*}_{-k}.
\label{eqn30}
\end{eqnarray}
 Detailed expressions for these terms are provided in Appendix.~\ref{AppendixA}. Therefore, $|\Psi_{0}\rangle$ and $\langle \Psi_{0}|$ can be expressed as:
\begin{subequations}
\begin{align}
|\Psi^{\gamma}_{0}\rangle &=\frac{1}{\mathcal{N}}\prod_{k>0}\Big(|\emptyset \rangle+i \alpha_{k}|k,-k\rangle \Big),\label{state_0_right}\\
 \langle\Psi^{\widetilde{\gamma}}_{0}|&=\frac{1}{\mathcal{N}}\prod_{k>0}\Big(\langle\widetilde{\emptyset}|-i \alpha^{*}_{k}\langle \widetilde{k},-\widetilde{k}|\Big)\label{state_0_left}.
\end{align}
\end{subequations}
Here, we defined $\alpha_{k}=Y_{k}/X_{k}$, with the state $|k,-k\rangle$ constructed as $\hat{\gamma}^{*}_{k}\hat{\gamma}^{*}_{-k}|\emptyset\rangle$ and the corresponding dual state $\langle \widetilde{k},-\widetilde{k}|=  \langle\widetilde{\emptyset}|\hat{\widetilde{\gamma}}_{-k}\hat{\widetilde{\gamma}}_{k}$. The normalization constant $\mathcal{N}$ is determined by evaluating $\langle \widetilde{\emptyset}|\emptyset\rangle$ and  $\langle \widetilde{k},-\widetilde{k}|k,-k\rangle$, based on the relationship between $\hat{\gamma}_{k}$ and $\hat{\widetilde{\gamma}}_k$. Now, by applying the evolution operators 
$e^{-i H_{\rm eff} t}$ and $e^{i H^{\dagger}_{\rm eff} t}$ as outlined in Eq.~\eqref{G-function}, we obtain the characteristic function in the $k$-basis as follows:
\begin{eqnarray}
\mathcal{G}_{\{0\}}= \frac{\langle \Psi^{\widetilde{\gamma}}_{0}(t)|e^{-i H_{\rm eff} u}|\Psi^{\gamma}_{0}(t)\rangle e^{i E^{i}_{0}u}}{\langle\Psi^{\widetilde{\gamma}}_{0}(t)|\Psi^{\gamma}_{0}(t)\rangle},
\label{eqn9}
\end{eqnarray}
where, 
\begin{subequations}
\begin{align}
|\Psi^{\gamma}_{0}(t)\rangle &= \frac{e^{-i E^{\rm eff}_{0}t}}{\mathcal{N}}\prod_{k>0}\Big(|\emptyset \rangle\!+\!i \alpha_{k}e^{-2i \epsilon^{\rm eff}_k t}|k,-k\rangle\Big)\label{stateright}\\
\langle\Psi^{\widetilde{\gamma}}_{0}(t)| &= \frac{e^{i E^{\rm eff^*}_{0}t}}{\mathcal{N}}\prod_{k>0}\Big(\langle\widetilde{\emptyset}|-i \alpha^{*}_{k}e^{2i \epsilon^{\rm eff^*}_kt}\langle \widetilde{k},-\widetilde{k}|\Big)\label{stateleft}
\end{align}
\end{subequations}\\
Below for concreteness, we focus on determining both the mean and the fluctuations of the work, 
$\langle W \rangle=-i\partial_u \log(\mathcal{G}_{\{0\}})$ and $\langle \Delta W^2\rangle=-\partial^2_u \log(\mathcal{G}_{\{0\}})$ respectively, from the following relations:
\begin{eqnarray}
&\langle W \rangle = -E_0^i+\frac{\langle \Psi^{\widetilde{\gamma}}_{0}(t)| H_f|\Psi^{\gamma}_{0}(t)\rangle}{\langle\Psi^{\widetilde{\gamma}}_{0}(t)|\Psi^{\gamma}_{0}(t)\rangle},\label{avg_work}\\
&\langle \Delta W^2\rangle 
=\frac{\langle \Psi^{\widetilde{\gamma}}_{0}(t)|(H_f)^2|\Psi^{\gamma}_{0}(t)\rangle}{\langle\Psi^{\widetilde{\gamma}}_{0}(t)|\Psi^{\gamma}_{0}(t)\rangle}\!-\!\Bigg(\!\frac{\langle \Psi^{\widetilde{\gamma}}_{0}(t)|H_f|\Psi^{\gamma}_{0}(t)\rangle}{\langle\Psi^{\widetilde{\gamma}}_{0}(t)|\Psi^{\gamma}_{0}(t)\rangle}\Bigg)^2
\label{fluctuations}
\end{eqnarray}
Since both Eq.~(\ref{stateright}) and (\ref{stateleft}) are expressed in terms of the left-right quasiparticle states of the system, it is clear that we should re-express $H_f$ in terms of $\hat{\gamma}_k$ and $\hat{\widetilde{\gamma}}_k$ as well, paying particular attention to maintaining the explicit Hermiticity of $H_f$.  While $H_{f}$ is itself Hermitian, expressing it solely through $\hat{\gamma}_{k}$ or $\hat{\widetilde{\gamma}}_{k}$ introduces non-Hermiticity. To address this, we employ a symmetrized version: $H_{f}(\gamma_{k},\widetilde{\gamma}_{k})=\{H_{f}(\gamma_{k})+H^{\dagger}_{f}(\widetilde{\gamma}_{k})\}/2$, which is Hermitian by construction. Detailed expressions can be found in Appendix.~\ref{AppendixA}.\\~\\
For trajectories within the transient (finite $t$) Fig.~\ref{Fig1}(a,b) illustrates how the average work density $\langle w \rangle$ (with $w=W/L$) and its variance $\langle \Delta w^2 \rangle$ change as the measurement strength $\gamma$ increases. When $\gamma=0$, both $\langle w \rangle$ and $\langle \Delta w^2 \rangle$ are zero because the Hamiltonians $H_{f}$ and $H_{\rm eff}$ coincide. 
However, as $\gamma$ increases, the non-Hermitian dynamics gradually projects the system onto the \textit{no-click} (spin-down) state. This projection leads to a reduction in fluctuations and the saturation of the work performed on the system, since further measurements cannot add energy once the system is sufficiently aligned with the \textit{no-click} state. This behavior is analogous to the quantum Zeno effect, where frequent “observations” effectively confine the system to a specific subspace. \\~\\
If instead we focus on the time-independent steady-state analysis of this model~\cite{turkeshi_4}, we observe signatures of the transition from the logarithmically entangled phase to the area law phase, which is set by the opening of an imaginary gap in the complex spectrum of the work statistics. 
Such stationary state phase transition turns out to be accompanied by a change in the structure of both the stationary state and the non-Hermitian vacuum~\cite{silva_2}, with the corresponding spectrum evolving from a phase with gapless $\Gamma_k$ to one with a finite gap~\cite{turkeshi_5}. Interestingly, as shown in Fig.~\ref{Fig2}, a feature in the average work $\langle w \rangle$, corresponding to a sharp kink near the critical  $\gamma_{c}(h)=4\sqrt{1-h^2}$ for each value of $h$, is clearly observed at sufficiently long times ($\Gamma_kt\gg 1, \forall k$). For larger $\gamma$, all modes have sizable decay rates, coherent evolution is Zeno-suppressed, and the available work slowly saturates.\\
%%%%%%%%%%%%%%%%Figure 3%%%%%%%%%%%%%%%%%%%%%%
\begin{figure}[h]
\centering
\includegraphics[width=0.8\columnwidth]{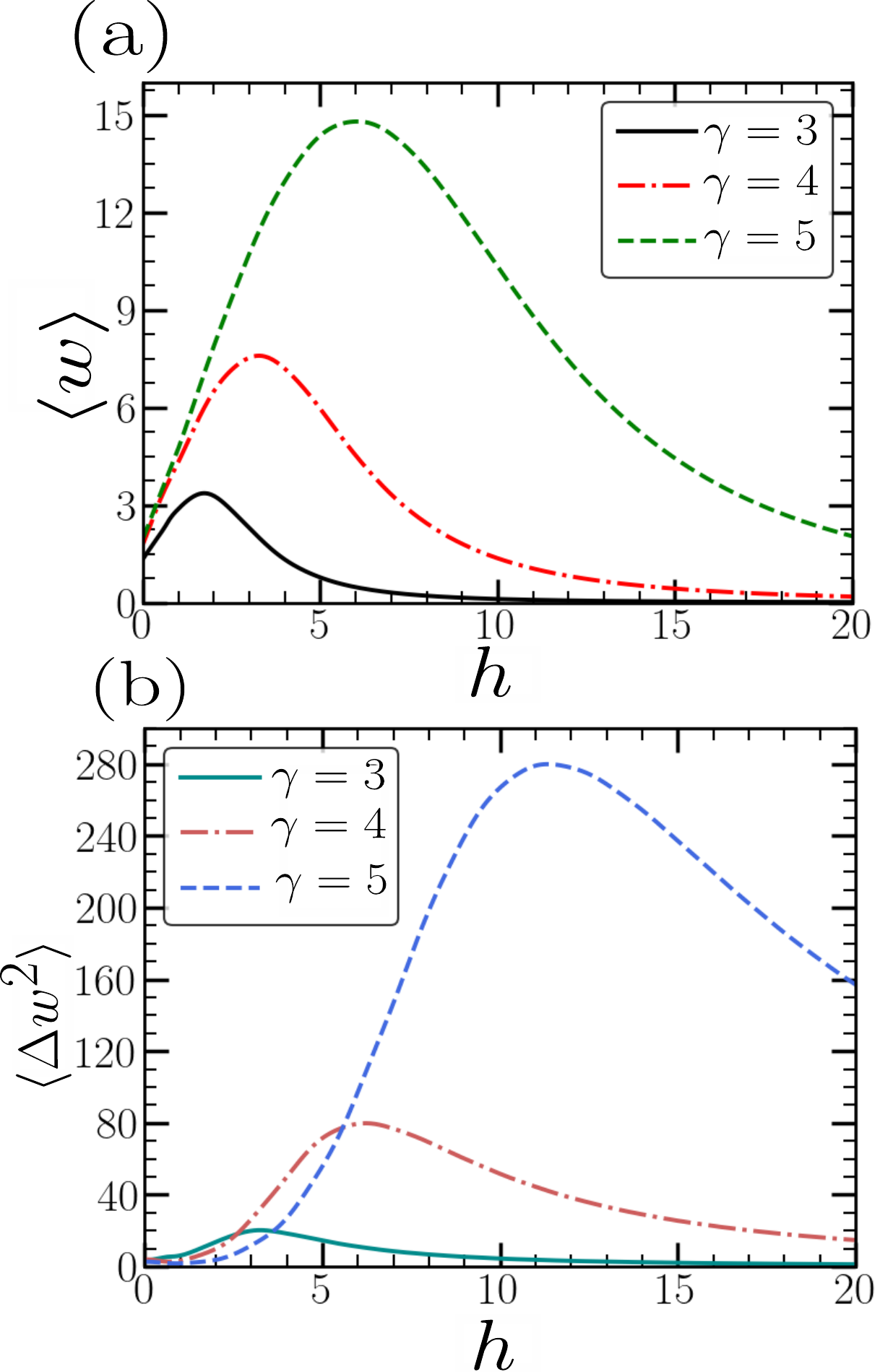}
\caption{(a) Average work density $\langle w \rangle$  and (b) its variance $\langle \Delta w^2 \rangle$ as functions of the transverse field $h$ for various fixed measurement strengths $\gamma$ at time $t=1.0$. The plots illustrate that for a given $\gamma$, both quantities initially increase with $h$, reaching a maximum before decreasing as $h$ is further increased, reflecting the interplay between unitary dynamics and measurement backaction. At higher $\gamma$ values, a stronger transverse field is required to overcome the measurement-induced energy modifications.}
\label{Fig3}
\end{figure}
%%%%%%%%%%%%%%%%%%%%%%%%%%%%%%%%%%%%%%%%%%%%%%

On the other hand, for fixed values of $\gamma$,  increasing the transverse field $h$ leads to a regime in which the unitary dynamics dominate over the measurement-induced corrections. In this limit, both the final Hamiltonian $H_{f}$ and the non-Hermitian Hamiltonian $H_{\rm eff}$  are essentially governed by the transverse field term, so that they become nearly identical. This leads to the asymptotic vanishing of the average work. As illustrated in Fig.~\ref{Fig3}(a,b), both $\langle w \rangle$ and $\langle \Delta w^2 \rangle$ initially increase with the transverse field $h$, reaching a maximum at a value that depends on the measurement strength $\gamma$. Beyond this point, as $h$ is further increased, both the quantities gradually decrease. Notably, for higher values of $\gamma$, a stronger transverse field is required to counterbalance the measurement. In Appendix~\ref{AppendixB}, we also present a detailed calculation of the efficacy $\gamma_t$ and plot its time dependence $\gamma_t(t)$ for different measurement rates $\gamma$.

\section{Conclusions}
\label{conclusion}
In this work, we presented a comprehensive study of quantum work statistics under continuous monitoring, revealing fundamental modifications to standard fluctuation theorems, particularly in the \it no-click \rm limit of quantum trajectories. Using the notion of quantum trajectories in the context of quantum thermodynamics~\cite{Elouard17} in combination with the standard two-point measurement scheme, we discuss a general expression for the work generating function that is easily specialized to a specific \textit{no-click} record and its associated non-Hermitian evolution. Notably, we show that generalized measurements necessitate a modification of the Jarzynski equality to account for non-Hermitian dynamics.\\~\\
Equipped with this formulation, we applied our theory to a one-dimensional transverse-field Ising model under local transverse spin monitoring to investigate the quantum work distribution arising from the interplay between coherent unitary evolution and non-Hermitian dynamics. We observe  measurement-induced energy saturation and the suppression of work fluctuations, both originating from an effective confinement of quantum states reminiscent of the quantum Zeno effect. Specifically, we observed that increasing the measurement strength $\gamma$ progressively projects the system onto the \textit{no-click} eigenstate, thus altering energy-exchange processes. In contrast, increasing the transverse field $h$ initially amplified energy fluctuations before eventually diminishing them, indicating a transition toward a regime dominated by unitary dynamics. Our results also show that the average work and its fluctuations are governed by the same decay rates that drive the steady-state entanglement transition in the \textit{no-click} Ising chain, thus providing a practical and experimentally accessible way to pinpoint the critical measurement rate for a given field using energy-based protocols rather than full state tomography. On current hardware with mid-circuit readout, including superconducting-qubit processors and trapped-ion systems, these observables can be measured during continuous monitoring, making them natural tools for benchmarking monitored many-body dynamics~\cite{Bruzewicz2019,Gaebler2022,Bluvstein2022,Kim2023}. Moreover, these underlying mechanisms also enable Zeno-based control and sensing, where tuning the measurement strength suppresses coherent dynamics in a controlled manner~\cite{Haase2024,Endo2021}. Taken together, this work advances both the theoretical framework for quantum work statistics in open systems undergoing continuous and generalized measurements and its concrete application to a many-body \textit{no-click} setting. Future research can extend these ideas to broader classes of quantum systems and alternative measurement protocols to further clarify implications and expand practical applications.

\section*{Acknowledgements}
A. S. and M. M acknowledge the support of the grant PNRR MUR project PE0000023-NQSTI. A. S. acknowledges support of the project “Superconducting quantum-classical linked computing systems (SuperLink)”, in the frame of QuantERA2 ERANET COFUND in Quantum Technologies.

\appendix

\section{Detailed Derivations for the Monitored Quantum Ising Chain}
\label{AppendixA}

Here, we provide a comprehensive derivation of the expressions used in Sec.~\ref{ising result}. We begin by revisiting  Eq.~(\ref{H_NC}) for $\gamma=0$, which after Jordan-Wigner transformation, can be expressed in terms of free fermions as follows:

\begin{eqnarray}
H_{i} =&-&J\sum_{i}\big(\hat{c}_{i}^{\dagger}\hat{c}_{i+1}+ \hat{c}_{i}^{\dagger}\hat{c}^{\dagger}_{i+1} + h.c\big) \notag\\
&-&h\big(1-2\hat{c}_{i}^{\dagger}\hat{c}_{i}\big)  
\end{eqnarray}
These operators satisfy the fermionic anticommutation relations: $\{\hat{c}_i,\hat{c}^{\dagger}_j\}=\delta_{ij}$ and $\{\hat{c}_i,\hat{c}_j\}=\{\hat{c}^{\dagger}_i,\hat{c}^{\dagger}_j\}=0$. Followed by the Fourier transformations of $\hat{c}_{i}$: $\hat{c}_{k}=\sum_{R_{i}}e^{-i k R_{i}}\hat{c}_{i}/\sqrt{L}$, we obtain: 

\begin{eqnarray}
H_{i} =  \sum_{k}
\begin{pmatrix}
\hat{c}^{\dagger}_{k} & \hat{c}_{-k}
\end{pmatrix}  
\begin{pmatrix}
a_{k} & b^{*}_{k}\\
b_{k} & -a_{k}
\end{pmatrix} \begin{pmatrix}
\hat{c}_{k}\\
\hat{c}^{\dagger}_{-k}
\end{pmatrix}
\label{A2}
\end{eqnarray}\\
where, $a_{k}=2(h-J\cos{k})$ and $b_{k}=2i J\sin{k}$. Now followed by the generalized Bogoliubov transformation, Eq.~(\ref{A2}) becomes:

\begin{eqnarray}
H_{i} &=&  \sum_{k}
\begin{pmatrix}
\hat{c}^{\dagger}_{k} & \hat{c}_{-k}
\end{pmatrix} \hat{V}^i_{k}\hat{V}^{i^{-1}}_{k} 
\begin{pmatrix}
a^i_{k} & b^{i*}_{k}\\
b^i_{k} & -a^i_{k}
\end{pmatrix} \hat{V}^i_{k}\hat{V}^{i^{-1}}_{k} \begin{pmatrix}
\hat{c}_{k}\\
\hat{c}^{\dagger}_{-k}
\end{pmatrix}
\notag\\
&=&\sum_{k} \begin{pmatrix}
\hat{\eta}^{\dagger}_{k} & \hat{\eta}_{-k}
\end{pmatrix} \begin{pmatrix}
\epsilon^i_{k} & 0\\
0 & -\epsilon^i_{k}
\end{pmatrix} \begin{pmatrix}
\hat{\eta}_{k}\\
\hat{\eta}^{\dagger}_{-k}
\end{pmatrix}
\label{A3}
\end{eqnarray}\\
As evident, Eq.~(\ref{A3}) is nothing but Eq.~(\ref{H_i}). Here, the eigenvector $\hat{V}^i_{k}$ is given by:
\begin{eqnarray}
V^i_{k} = \begin{pmatrix}
u^i_{k} & -i v^i_{k}\\
-i v^i_{k} & u^i_{k}
\end{pmatrix}
\label{A4}
\end{eqnarray}
with, $u^i_{k}=1\big/\sqrt{1+\big|\frac{b^i_k}{a^i_k+\epsilon^i_k}\big|^2}$ and $v^i_k=i b^i_k/(a^i_k+\epsilon^i_k). u^i_k$. The free fermions $\hat{c}_k$ are related to the quasiparticles $\hat{\eta}_{k}$ through the following Bogoliubov transformations:

\begin{eqnarray}
&\hat{c}_{k} = u^i_k \hat{\eta}_{k}  - i v^i_k \hat{\eta}^{\dagger}_{-k}\notag\\
&\hat{c}_{-k} = u^i_k \hat{\eta}_{-k}  + i v^i_k \hat{\eta}^{\dagger}_{k}\notag\\
&\hat{c}^{\dagger}_{k} = u^i_k \hat{\eta}^{\dagger}_{k}  + i v^i_k \hat{\eta}_{-k}\notag\\
&\hat{c}^{\dagger}_{-k}= u^i_k \hat{\eta}^{\dagger}_{-k}  - i v^i_k \hat{\eta}_{k}
\label{A5}
\end{eqnarray}

Similarly, Eq.~(\ref{H_nc_a}) (or Eq.~(\ref{H_nc_b}) for $H_{\rm eff}^{\dagger}$) can be obtained from,

\begin{align}
H_{\rm eff} &= \sum_{k}
\begin{pmatrix}
\hat{c}^{\dagger}_{k} & \hat{c}_{-k}
\end{pmatrix} V^{\rm nc}_{k}V^{{\rm nc}^{-1}}_{k} 
\begin{pmatrix}
a^{\rm nc}_{k} & b^{{\rm nc}*}_{k}\\
b^{\rm nc}_{k} & -a^{\rm nc}_{k}
\end{pmatrix} V^{\rm nc}_{k}V^{{\rm nc}^{-1}}_{k}\begin{pmatrix}
\hat{c}_{k}\\
\hat{c}^{\dagger}_{-k}
\end{pmatrix}
\notag\\
&= \sum_{k} \begin{pmatrix}
\hat{\gamma}^{*}_{k} & \hat{\gamma}_{-k}
\end{pmatrix} \begin{pmatrix}
\epsilon^{\rm eff}_{k} & 0\\
0 & -\epsilon^{\rm eff}_{k}
\end{pmatrix} \begin{pmatrix}
\hat{\gamma}_{k}\\
\hat{\gamma}^{*}_{-k}
\end{pmatrix}
\label{A6}
\end{align}

with the eigenvector $V^{\rm nc}_{k}$ taking the same form as in Eq.~(\ref{A4}), but with the superscript $``i"$ replaced by $``{\rm nc}"$. The elements of $V^{\rm nc}_{k}$ are $u^{\rm nc}_{k}=1\big/\sqrt{1+\big|\frac{b^{\rm nc}_k}{a^{\rm nc}_k+\epsilon^{\rm nc}_k}\big|^2}$ and $v^{\rm nc}_k=i b^{\rm nc}_k/(a^{\rm nc}_k+\epsilon^{\rm nc}_k). u^{\rm nc}_k$, with $a^{\rm nc}_{k}=2(h-J\cos{k})+i \gamma/2$ and $b^{\rm nc}_k=b^i_{k}$. Both the Hermitian and non-Hermitian quasiparticles, $\hat{\eta}_k$ and $\hat{\gamma}_k$, obey the similar anticommutation relations as fermions. From Eq.~(\ref{A6}), the Bogoliubov transformation connecting the non-Hermitian quasiparticles $\hat{\gamma}_{k}$ and fermions $\hat{c}_{k}$ can be written as: 

\begin{eqnarray}
&\hat{c}_{k} = u^{\rm nc}_k \hat{\gamma}_{k}  - i v^{\rm nc}_k \hat{\gamma}^{*}_{-k}\notag\\
&\hat{c}_{-k} = (u^{\rm nc}_k \hat{\gamma}_{-k}  + i v^{\rm nc}_k \hat{\gamma}^{*}_{k})/\det(V_{k}^{\rm nc})\notag\\
&\hat{c}^{\dagger}_{k} = (u^{\rm nc}_k \hat{\gamma}^{*}_{k}  + i v^{\rm nc}_k \hat{\gamma}_{-k})/\det(V_{k}^{\rm nc})\notag\\
&\hat{c}^{\dagger}_{-k}= u^{\rm nc}_k \hat{\gamma}^{*}_{-k}  - i v^{\rm nc}_k \hat{\gamma}_{k}
\label{A7}
\end{eqnarray}
Similarly, the mapping between \(\hat{c}_k\) and the left non‑Hermitian quasiparticles \(\hat{\widetilde\gamma}_k\) is obtained by taking the complex conjugate of each Bogoliubov coefficient in Eq.~(\ref{A7}) -- that is, \(u_k^{\rm nc}\to u_k^{\rm nc*}\) and \(v_k^{\rm nc}\to v_k^{\rm nc*}\), which defines the corresponding rotation matrix \({\widetilde V}_k^{\rm nc}\). Then, by comparing Eq.~(\ref{A5}) and Eq.~(\ref{A7}), we immediately derive the relation between $\hat{\eta}_{k}$ and $\hat{\gamma}_{k}$, as given by Eq.~(\ref{eqn30}) in the main text. Where,
\begin{eqnarray}
 X_k=u_{k}^i u^{\rm nc}_{k} +  v_{k}^i v^{\rm nc}_{k}\notag\\
 Y_k=u_{k}^i v^{\rm nc}_{k} -  v_{k}^i u^{\rm nc}_{k}.
 \end{eqnarray}
Likewise, replacing $X_k$ and $Y_k$ with their complex conjugates in that same equation produces the equivalent mapping between $\hat{\eta}_{k}$ and the left non-Hermitian quasiparticles $\hat{\widetilde{\gamma}}_{k}$.\\

We now derive the analytical expression for the symmetrized Hamiltonian $H_{f}(\gamma_k,\widetilde{\gamma}_k)$. We begin with the form of $H_f$ from Eq.~(\ref{H_i}) and then rewrite $\hat{\eta}_{k}$ in terms of $\hat{\gamma}_{k}$ by using the relations provided in Eq.~(\ref{A5}) and (\ref{A7}). Consequently, the final Hamiltonian $H_f(\gamma_k)$ can be written as follows:

\begin{eqnarray}
H_f(\gamma_k)=\sum_{k>0} \beta_{k}N_{k} -i \delta_{k}\Pi_{k} + E_k^{0\gamma}
\label{A8}
\end{eqnarray}
where the number operator is defined as $N_{k}=\hat{\gamma}^{*}_{k} \hat{\gamma}_k + \hat{\gamma}_{-k}^{*} \hat{\gamma}_{-k}$ and the pair operator as $\Pi_{k}=\hat{\gamma}^{*}_{k} \hat{\gamma}^{*}_{-k} + \hat{\gamma}_{k} \hat{\gamma}_{-k}$. The prefactors are given by: 
\begin{eqnarray}
\beta_k = \frac{\epsilon_k^i \left(X_k^2 - Y_k^2\right)}{\det\left(V_k^{\rm nc}\right)}\;\text{and}\;\delta_k = \frac{2\epsilon_k^i X_k Y_k}{\det\left(V_k^{\rm nc}\right)}
\label{A9}
\end{eqnarray}
\vspace{1em}
and the ground state energy after transforming from $\hat{\eta}_k$ to $\hat{\gamma}_k$ is expressed as: 
\begin{eqnarray}
E_k^{0\gamma}=-\sum_{k>0}\epsilon_{k}^{i}+\frac{2\epsilon^{i}_{k}Y_k^2}{\det(V_k^{\rm nc})}  
\label{A10}
\end{eqnarray}
Similarly, by applying the analogous procedure, we infer the expression for $H_f^{\dagger}(\widetilde{\gamma}_k)$ as below: 
\begin{eqnarray}
H_f^{\dagger}(\widetilde{\gamma}_k)=\sum_{k>0} \beta^*_{k}{\widetilde{N}}_{k} -i \delta^*_{k}{\widetilde{\Pi}}_{k} + E_k^{0\widetilde{\gamma}}
\label{A11}
\end{eqnarray}
Here, ${\widetilde{N}}_{k}$ and ${\widetilde{\Pi}}_{k}$ denote the number and pair operators for $\hat{\widetilde{\gamma}}_{k}$, respectively, and the pre-factors as well as the ground state energy in Eq.~(\ref{A11}) are given by the complex conjugates of those in Eq.~(\ref{A9}) and (\ref{A10}).\\

Now that we have the expressions for $H_{f}(\gamma_k)$ and $H_f^{\dagger}(\widetilde{\gamma}_k)$ in hand, we can proceed to formulate the average energy density $\langle w \rangle$ and its fluctuations $\langle \Delta w^2 \rangle$  by performing the appropriate averages over the initial states $|\Psi^{\gamma}_{0}(t) \rangle$ and $\langle\Psi^{\widetilde{\gamma}}_{0}(t)|$. After carrying out these calculations, Eq.~(\ref{avg_work}) takes the following form:
\begin{widetext}
\begin{eqnarray}
\langle w\rangle=-E_0^i+ \int^{2\pi}_{0}\frac{dk}{2\pi}\;\frac{2(\beta_k+\beta^{*}_k)|\alpha_k|^2e^{-4\Gamma_kt}-(\delta_k+\delta^{*}_k)(\alpha^t_k+\alpha^{t*}_k)+(E_k^{0\gamma}+E_k^{0\widetilde{\gamma}})(1+|\alpha_k|^2e^{-4\Gamma_kt})}{2(1+|\alpha_k|^2e^{-4\Gamma_kt})}
\label{A12}
\end{eqnarray}
\end{widetext}
here, $\alpha^t_{k}=\alpha_{k}e^{-2i \epsilon^{\rm eff}_{k}t}$, and $\alpha^{t*}_{k}$ denotes its complex conjugate. However, evaluating the first term in Eq.~(\ref{fluctuations}) is nontrivial because the square $\big(H_{f}(\gamma_{k})+H_f^{\dagger}(\widetilde{\gamma}_{k})\big)^2$ expands into four different contributions: $[H_{f}(\gamma_{k})]^2+[H_f^{\dagger}(\widetilde{\gamma}_{k})]^2+H_{f}(\gamma_{k}).H_f^{\dagger}(\widetilde{\gamma}_{k})+H_f^{\dagger}(\widetilde{\gamma}_{k}).H_{f}(\gamma_{k})$. Of these four contributions, the mixed term, $H_f(\gamma_{k}).H_f^{\dagger}(\widetilde{\gamma}_{k})$ requires additional treatment. To evaluate the expectation value $\langle \Psi^{\widetilde{\gamma}}_{0}(t)|H_{f}(\gamma_{k}).H_f^{\dagger}(\widetilde{\gamma}_{k})|\Psi^{\gamma}_{0}(t)\rangle$ properly, the operator $H_{f}(\gamma_{k})$ acting on the left state must be rewritten in the $\widetilde{\gamma}_k$-basis, while the operator $H_f^{\dagger}(\widetilde{\gamma}_{k})$ acting on right state must be expressed in the $\gamma_k$-basis. To accomplish this, we construct the following Bogoliubov-like rotation linking $\hat{\gamma}_k$ and $\hat{\widetilde{\gamma}}_k$ by combining the transformation between $\hat{c}_k$ and $\hat{\gamma}_k$ (Eq.~(\ref{A7})) with the analogous relation between $\hat{c}_k$ and $\hat{\widetilde{\gamma}}_k$:
\begin{eqnarray}
\hat{\gamma}_k=(P_k \hat{\widetilde{\gamma}}_k +i Q_k\hat{\widetilde{\gamma}}^*_{-k})/\det(V^{\rm nc}_{k})\notag\\
\hat{\gamma}_{-k}=(P_k \hat{\widetilde{\gamma}}_{-k} -i Q_k\hat{\widetilde{\gamma}}^*_{k})/\det(\widetilde{V}^{\rm nc}_{k})\notag\\
\hat{\gamma}^{*}_{k}=(P_k \hat{\widetilde{\gamma}}^{*}_{k} -i Q_k\hat{\widetilde{\gamma}}_{-k})/\det(\widetilde{V}^{\rm nc}_{k})\notag\\
\hat{\gamma}^{*}_{-k}=(P_k \hat{\widetilde{\gamma}}^*_{-k} +i Q_k\hat{\widetilde{\gamma}}_{k})/\det(V^{\rm nc}_{k})
\label{A13}
\end{eqnarray}\\
where $P_k=|u_k^{\rm nc}|^2+|v_k^{\rm nc}|^2$ and $Q_k=u_k^{\rm nc*}v_{k}^{\rm nc}-u_k^{\rm nc}v_{k}^{\rm nc*}$.\\

We are now ready to evaluate the expectation value in the first term of Eq.~(\ref{fluctuations}). This term can be decomposed into three distinct contributions, as follows:

\begin{widetext}
\begin{eqnarray}
&\text{First term;}\quad\langle \Psi^{\widetilde{\gamma}}_{0}(t)|[H_f(\gamma_k)]^2+[H_f^{\dagger}(\widetilde{\gamma}_k)]^2|\Psi_0^{\gamma}(t)\rangle= \notag\\[3ex]
&\genfrac{}{}{0.9pt}{0}{
  \begin{aligned}
    & 4|\alpha_k|^2 e^{-4\Gamma_kt}(\beta_k^2+\beta_k^{*2})+(\delta_k^2+\delta_k^{*2})(1+|\alpha_k|^2e^{-4\Gamma_kt})- 2(\alpha^t_k+\alpha^{t*}_k)\big[\delta_k(\beta_k+E^{0\gamma}_{k})+\delta^*_k(\beta^*_k+E^{0\widetilde{\gamma}}_{k})\big] \\
    &+ (E^{0\gamma}_{k}+E^{0\widetilde{\gamma}}_{k})
    + |\alpha_k|^2e^{-4\Gamma_kt}\big[E^{0\gamma}_{k}(4\beta_k+E^{0\gamma}_{k})+E^{0\widetilde{\gamma}}_{k}(4\beta^*_k+E^{0\widetilde{\gamma}}_{k})\big]
  \end{aligned}
}{
  \begin{aligned}
    4(1+|\alpha_k|^2 e^{-4\Gamma_kt})
  \end{aligned}
}\notag\\[6ex]
&\text{Second term;}\quad\langle \Psi^{\widetilde{\gamma}}_{0}(t)|H_f^{\dagger}(\widetilde{\gamma}_k).H_f(\gamma_k)|\Psi_0^{\gamma}(t)\rangle= \notag\\[3ex]
&\genfrac{}{}{0.9pt}{0}{
  \begin{aligned}
    & 4|\alpha_k|^2 |\beta_k|^2  e^{-4\Gamma_kt}-2(\alpha^{t*}_k\beta^*_k\delta_k+\alpha^{t}_k\beta_k\delta^*_k)+2|\alpha_k|^2 e^{-4\Gamma_kt} (\beta_k E^{0\widetilde{\gamma}}_{k}+\beta^*_k E_k^{0\gamma})+|\delta_k|^2(1+e^{-4\Gamma_kt}) \\
    &-(E_k^{0\widetilde{\gamma}}\delta_k+E_k^{0\gamma}\delta^*_k)(\alpha_k^t+\alpha_k^{t*})+E_k^{0\gamma}E_k^{0\widetilde{\gamma}}(1+|\alpha_k|^2e^{-4\Gamma_kt})
  \end{aligned}
}{
  \begin{aligned}
    4(1+|\alpha_k|^2 e^{-4\Gamma_kt})
  \end{aligned}
}\notag\\[6ex]
&\text{Third term;}\quad\langle \Psi^{\widetilde{\gamma}}_{0}(t)|H_f(\gamma_k).H_f^{\dagger}(\widetilde{\gamma}_k)|\Psi_0^{\gamma}(t)\rangle= \notag\\[3ex]
&\genfrac{}{}{0.9pt}{0}{
  \begin{aligned}
    & A\widetilde{A}+B\widetilde{B} 
  \end{aligned}
}{
  \begin{aligned}
    4(1+|\alpha_k|^2 e^{-4\Gamma_kt})
  \end{aligned}
},\quad \text{where}\notag\\[6ex]
&A=\Big(2i \alpha^t_k\big[(P_k^2-Q_k^2)\beta^*_k-2\delta^*_kP_kQ_k\big]-i \delta^*_k(P_k^2-Q_k^2)-2i \beta^*_kP_kQ_k+2i \alpha^t_k\big[\beta^*_kQ^2_k
+\delta^*_kP_kQ_k\big]+i \alpha^t_k E^{0\widetilde{\gamma}}_k\Big)\big/D_k\notag\\[1ex]
&\widetilde{A}=\Big(-2i \alpha^{t*}_k\big[(P_k^2-Q_k^2)\beta_k+2\delta_kP_kQ_k\big]+i \delta_k(P_k^2-Q_k^2)-2i \beta_kP_kQ_k-2i \alpha^{t*}_k\big[\beta_kQ^2_k
-\delta_kP_kQ_k\big]-i \alpha^{t*}_k E^{0\gamma}_k\Big)\big/D_k\notag\\[1ex]
&B=E^{0\widetilde{\gamma}}_k+\genfrac{}{}{0.9pt}{0}{2}{D_k}\big[Q_k^2\beta^*_k+\delta^*_kP_kQ_k\big]-\genfrac{}{}{0.9pt}{0}{\alpha_k}{D_k}\big[\delta^*_k(P_k^2-Q_k^2)+2\beta^*_kP_kQ_k\big]\notag\\
&\widetilde{B}=E^{0\gamma}_k+\genfrac{}{}{0.9pt}{0}{2}{D_k}\big[Q_k^2\beta_k-\delta_kP_kQ_k\big]-\genfrac{}{}{0.9pt}{0}{\alpha^*_k}{D_k}\big[\delta_k(P_k^2-Q_k^2)-2\beta_kP_kQ_k\big]
\label{A14}
\end{eqnarray}
\end{widetext}
where, $D_k=\det(V^{\rm nc}_k\widetilde{V}^{\rm nc}_k)$. Now, by integrating the three terms in Eq.~(\ref{A14}) over $dk$ as prescribed, we evaluate the first term in Eq.~(\ref{fluctuations}), while, the second term is simply the square of the corresponding term in Eq.~(\ref{A12}).

\section{Calculation of the efficacy for Monitored Quantum Ising Chain}
\label{AppendixB}
Here we explicitly compute the efficacy $\gamma_t$ for the monitored quantum Ising chain. Setting $H_i=H_f$, Eq.~\eqref{efficacy} reduces to 
\begin{equation}
\gamma_t=\frac{\langle \Psi^{\widetilde{\gamma}}_0| e^{-i H_{{\rm eff}} t} e^{i H^\dagger_{{\rm eff}}t} |\Psi^{\gamma}_{0}\rangle}   {\langle \Psi^{\widetilde{\gamma}}_0| e^{i H^\dagger_{{\rm eff}} t} e^{-i H_{{\rm eff}}t}|\Psi^{\gamma}_{0}\rangle}.
\end{equation}
The denominator is just the biorthogonal overlap of the time-evolved left and right eigenstates in Eq.~(\ref{stateright})-(\ref{stateleft}). The numerator, in contrast, requires an explicit change of basis: using the rotation $\gamma_k \leftrightarrow \widetilde{\gamma}_k$ from Eq.~(\ref{A14}), we rewrite $H_{\rm eff}$ in terms of $\widetilde{\gamma}_k$ and $H^{\dagger}_{\rm eff}$ in terms of $\gamma_k$. With this choice, $H^{\dagger}_{\rm eff}$ acts on the right sector $|\Psi^{\gamma}_{0}\rangle$ and $H_{\rm eff}$ acts on the left sector $\langle \Psi^{\widetilde{\gamma}}_0|$ in their natural biorthogonal bases. Applying the basis transformation in Eq.~(\ref{H_nc_a})-(\ref{H_nc_b}), the Hamiltonians takes the form:
\begin{subequations}
\begin{align}
H_{\rm eff}(\widetilde{\gamma}) = \sum_{k>0}\big[\chi_{k} \widetilde{N}_k - i \xi_{k} \widetilde{\Pi}_k\big] + E_{0},\\
H_{\rm eff}^{\dagger}(\gamma) = \sum_{k>0}\big[\overline{\chi}_{k} N_k - i \overline{\xi}_{k} \Pi_k\big] + E_{0}^{*}.
\end{align}
\end{subequations}
Here the coefficients are 
\begin{equation}
\chi_k =  \frac{\epsilon_k^{\rm eff} \left(P_k^2 - Q_k^2\right)}{D_k} \quad \text{and}\quad \xi_k = - \frac{2\epsilon_k^{\rm eff} P_k Q_k}{D_k},
\label{A9}
\end{equation}
and the ground-state energy after the transformation is
\begin{equation}
E_0 = E_{0}^{\rm eff} + \sum_{k>0} \frac{2\epsilon_{k}^{\rm eff}Q_{k}^{2}}{D_k}.
\end{equation}
Here, $D_k = \det(V_k^\text{nc}) \det(\widetilde{V}_k^\text{nc})$, $\overline{\chi}_k=\chi_k^{*}$ and $\overline{\xi}_k=-\xi_k^{*}$.\\~\\
Let us first calculate the term $e^{i H_\text{eff}^\dagger t} |\Psi_0^\gamma \rangle$ in the numerator. Since $H_\text{eff}^\dagger = \sum_{k>0} h_k^\dagger + E_0$, is a sum of independent $k$-blocks with $[h_q^\dagger, h_k^\dagger] = 0$ for $q \neq k$, the evolution operator factorizes: $e^{iH^{\dagger}_{\rm eff}t}=e^{iE_0 t}\prod_{k>0} e^{ih^{\dagger}_{k}t}$. For each $k$, we work in the even-parity subspace spanned by $|0\rangle_k \equiv|\emptyset \rangle$ and $|1\rangle_k \equiv |k,-k\rangle$. Using the canonical anticommutation relations one checks: $N_k|0\rangle_k = 0$, $N_k|1\rangle_k = 2|1\rangle_k$, $\Pi_k|0\rangle_k = |1\rangle_k$, and $\Pi_k|1\rangle_k = -|0\rangle_k$. In the ordered basis $\{|0\rangle_k, |1\rangle_k\}$ these read:
\begin{equation}
N_{k} = \begin{pmatrix}
0 & 0\\
0 & 2
\end{pmatrix}  =\mathbb{1}-\sigma_z; \; \text{and} \;\Pi_{k} = \begin{pmatrix}
0 & -1\\
1 & 0
\end{pmatrix}=-i\sigma_y.
\end{equation}
Here $\sigma_{x,y,z}$ are the Pauli matrices acting in the $\{|0\rangle_k, |1\rangle_k\}$ space. The single-block Hamiltonian therefore can be written as: $h^{\dagger}_k=\overline{\chi}_k \mathbb{1}-(\overline{\chi}_k \sigma_z+\overline{\xi}_k \sigma_y)$. Using the identity for Pauli–vector exponentials, $e^{i\theta\mathbf{b}.\boldsymbol{\sigma}}=\cos{(|\mathbf{b}|\theta)}.\mathbb{1}+i\frac{\sin{(|\mathbf{b}|\theta)}}{|\mathbf{b}|}(\mathbf{b}.\boldsymbol{\sigma})$, with $\mathbf{b}_k=(0,-\overline{\chi}_k, -\overline{\xi}_k)$ and $\overline{\Omega}_k \equiv |\mathbf{b}_k|=\sqrt{\overline{\chi}_k^2+\overline{\xi}_k^2}$, we obtain:
\begin{equation}
 e^{ih_k^{\dagger}t}=e^{i\overline{\chi}_k t}\bigg[\cos{(\overline{\Omega}_k t)}.\mathbb{1}-i\frac{\sin{(\overline{\Omega}_k t)}}{\overline{\Omega}_k} \big(\overline{\chi}_k \sigma_z+\overline{\xi}_k \sigma_y \big)\bigg].\notag
\end{equation}
Equivalently, in the basis $\{|0\rangle_k,|1\rangle_k \}$, the above equation can be expressed as a $2\times2$ matrix:
%\frac{\overline{\chi}_k}{\overline{\Omega}_k}\big)
\begin{widetext}
\begin{equation}
e^{ih_k^{\dagger}t}=e^{i\overline{\chi}_k t} \begin{pmatrix}
\cos{(\overline{\Omega}_k t)}-i\big(\overline{\chi}_k/\overline{\Omega}_k\big)\sin{(\overline{\Omega}_k t)} & -\big(\overline{\xi}_k/\overline{\Omega}_k\big)\sin{(\overline{\Omega}_k t)} \\
\big(\overline{\xi}_k/\overline{\Omega}_k\big)\sin{(\overline{\Omega}_k t)} & \cos{(\overline{\Omega}_k t)}+i\big(\overline{\chi}_k/\overline{\Omega}_k\big)\sin{(\overline{\Omega}_k t)}
\end{pmatrix}   
\end{equation}
\end{widetext}
Acting on the two-component amplitude $C_k(0)=(1, i\alpha_k)^{\rm T}$ of the initial state $|\Psi^{\gamma}_0\rangle$ [Eq.~(\ref{state_0_right})] gives
\begin{equation}
\begin{pmatrix}
C_{0,k}(t)\\
C_{1,k}(t)
\end{pmatrix}=e^{ih_k^{\dagger}t} \begin{pmatrix}
1\\
i\alpha_k
\end{pmatrix},
\end{equation}
that is,
\begin{subequations}
\begin{align}
C_{0,k}(t) &= e^{i\overline{\chi}_k t}\bigg[\cos{(\overline{\Omega}_k t)} -i \frac{\overline{\chi}_k+\alpha_k \overline{\xi}_k}{\overline{\Omega}_k} \sin{(\overline{\Omega}_k t)}\bigg] \notag\\
C_{1,k}(t) &= e^{i\overline{\chi}_k t}\bigg[i\alpha_k\cos{(\overline{\Omega}_k t)} + \frac{\overline{\xi}_k-\alpha_k \overline{\chi}_k}{\overline{\Omega}_k} \sin{(\overline{\Omega}_k t)}\bigg]\notag.
\end{align}
\end{subequations}
Therefore, the final expression for $e^{i H_\text{eff}^\dagger t} |\Psi_0^\gamma \rangle$ is given by: 
\begin{equation}
e^{i H_\text{eff}^\dagger t} |\Psi_0^\gamma \rangle = \frac{e^{iE_{0}^{*}t}}{\mathcal{N}}\prod_{k} \big[C_{0,k}(t)|0\rangle_k + C_{1,k}(t)|1\rangle_k\big].
\end{equation}
Let us now compute the backward evolution term $\langle \Psi^{\widetilde{\gamma}}_{0}|e^{-iH_{\rm eff}t}$. To evaluate this, we work in the biorthogonal basis $\{\langle \widetilde{0}|_k,\langle \widetilde{1}|_k\}$ defined by $\langle \widetilde{0}|_k \equiv \langle \widetilde{\emptyset}|$ and $\langle \widetilde{1}|_k \equiv \langle \widetilde{k},-\widetilde{k}| = \langle \widetilde{\emptyset}|\widetilde{\gamma}_{-k} \widetilde{\gamma}_{k}$. In this basis, the dual number operator $\widetilde{N}_k$ and the dual pair operator $\widetilde{\Pi}_k$ act on the even sector as: $\widetilde{N}_k=\mathbb{1}-\sigma_z$ and  $\widetilde{\Pi}_k=i\sigma_y$. The $+i\sigma_y$ sign reflects the fact that we are acting on left (bra) states of the biorthogonal pair. With these $2\times 2$ blocks in hand, each $k$-sector can be exponentiated using the standard Pauli–matrix identity, and following the same steps as above we obtain the explicit form of $e^{-ih_k t}$ in the dual basis: 
\begin{widetext}
\begin{equation}
e^{-ih_k t}=e^{-i\chi_k t} \begin{pmatrix}
\cos{(\Omega_k t)}+i\big(\chi_k/\Omega_k\big)\sin{(\Omega_k t)} & -\big(\xi_k/\Omega_k\big)\sin{(\Omega_k t)} \\
\big(\xi_k/\Omega_k\big)\sin{(\Omega_k t)} & \cos{(\Omega_k t)}-i\big(\chi_k/\Omega_k\big)\sin{(\Omega_k t)}
\end{pmatrix}   
\end{equation}
\end{widetext}
Now acting on the initial amplitude row vector $D_k(0)=(1, -i\alpha_{k}^{*})$ of the left initial state $\langle \Psi^{\widetilde{\gamma}}_{0}|$ yields
\begin{equation}
 \begin{pmatrix}
D_{0,k}(t) &  D_{1,k}(t)
\end{pmatrix}= \begin{pmatrix}
1 & -i\alpha_k^{*}
\end{pmatrix} e^{-ih_k t},   
\end{equation}
where
\begin{subequations}
\begin{align}
D_{0,k}(t) &= e^{-i\chi_k t}\bigg[\cos{(\Omega_k t)} +i \frac{\chi_k-\alpha^{*}_k \xi_k}{\Omega_k} \sin{(\Omega_k t)}\bigg] \notag\\
D_{1,k}(t) &= e^{-i\chi_k t}\bigg[-i\alpha^{*}_k\cos{(\Omega_k t)} - \frac{\xi_k+\alpha^{*}_k \chi_k}{\Omega_k} \sin{(\Omega_k t)}\bigg]\notag.
\end{align}
\end{subequations}
Therefore, the final expression for the backward-evolved state is 
\begin{equation}
\langle \Psi^{\widetilde{\gamma}}_{0}|e^{-iH_{\rm eff}t} = \frac{e^{-iE_{0}t}}{\mathcal{N}}\prod_{k} \big[\langle \widetilde{0}_k|D_{0,k}(t) + \langle \tilde{1}_k|D_{1,k}(t)\big].
\end{equation}
Since by construction one has $D_{0,k}(t)=C^{*}_{0,k}(t)$ and $D_{1,k}(t)=C^{*}_{1,k}(t)$, the mode-resolved contribution to the efficacy is
\begin{equation}
\gamma^{(k)}_t = e^{-4Q^{2}_{k}\Gamma_k t/D_k}\Bigg(\frac{|C_{0,k}|^{2}+|C_{1,k}|^{2}}{1+|\alpha_{k}|^2 e^{-4\Gamma_{k}t}}\Bigg).
\end{equation}
The many-body efficacy is then the product over positive momenta: $\gamma_t =\prod_{k>0}\gamma^{(k)}_t$.
%%%%%%%%%%%%%%%%Figure 4%%%%%%%%%%%%%%%%%%%%%%
\begin{figure}[H]
\centering
\includegraphics[width=0.85\columnwidth]{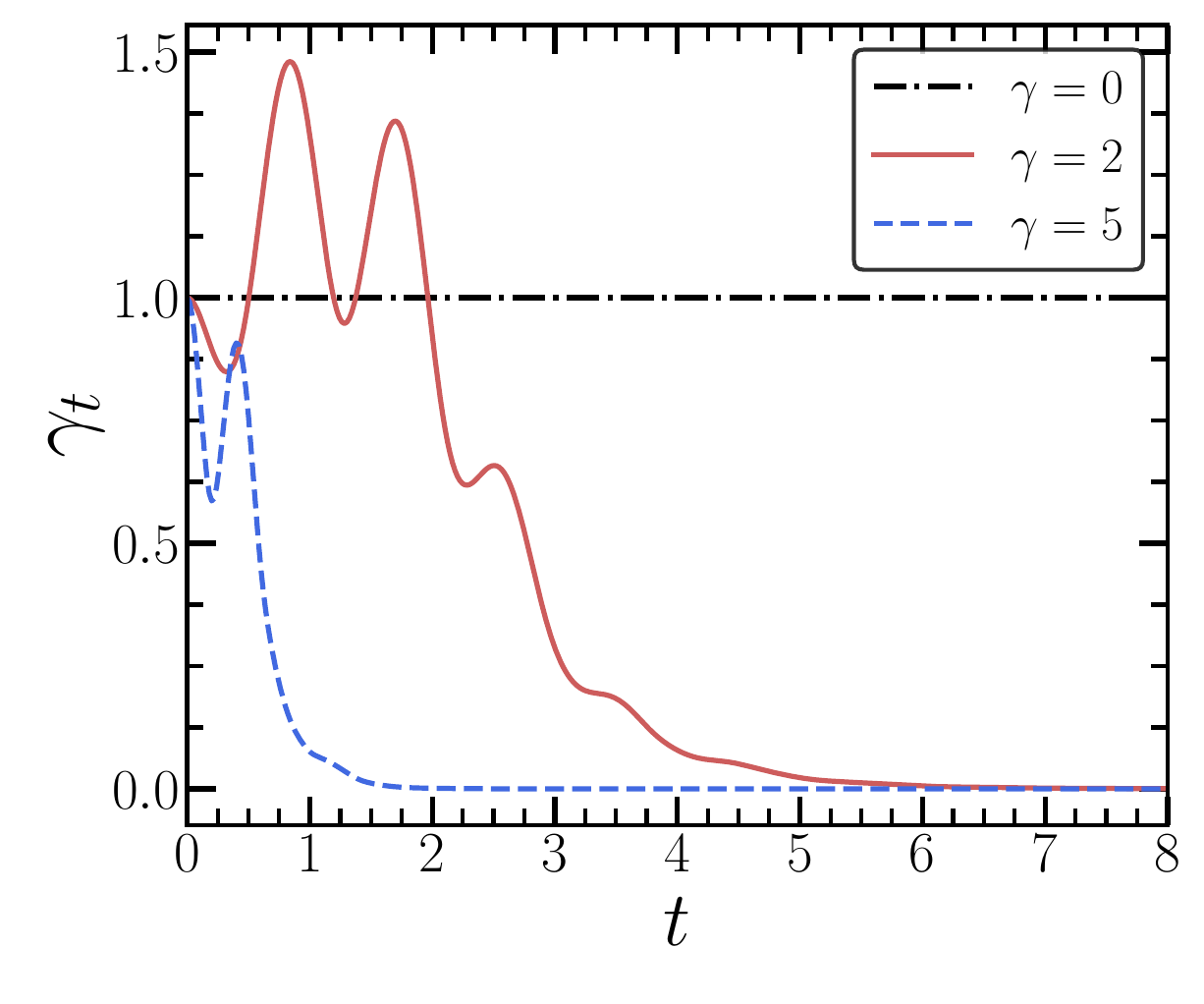}
\caption{Efficacy $\gamma_t$ versus time $t$ for the monitored Ising chain at $J=1$ and $h=0.5$, comparing three measurement rates. The unitary case $\gamma=0$ yields $\gamma_t=1$ for all times. For moderate monitoring at $\gamma=2$ the curve shows a brief overshoot at initial times then decays as the non-Hermitian loss builds up. For stronger monitoring, at $\gamma=5$, the decay is faster and the initial bump is much smaller.}
\label{Fig4}
\end{figure}
%%%%%%%%%%%%%%%%%%%%%%%%%%%%%%%%%%%%%%%%%%%%%%
Fig.~\ref{Fig4} illustrates the time dependence of the efficacy $\gamma_t$ at various measurement rates. As evident from the plot, in the absence of monitoring ($\gamma=0$), the evolution remains unitary, allowing the forward and backward segments to commute. Consequently, $\gamma_t$ remains constant at 1 for all times. However, any nonzero measurement strength $\gamma>0$ disrupts this symmetry, , rendering the post-selected \textit{no-click} evolution effectively non-Hermitian. Each Bogoliubov mode then acquires a complex quasi-energy $\epsilon_{k}^{\rm eff}$, where the real part induces coherent oscillations, and the imaginary part drives the decay of the no-jump weight. For $\gamma=2$, small oscillations appear at short times when the loss is moderate. As time progresses, accumulated decay rates across modes begin to dominate, causing $\gamma_t$ to drop towards zero. For strong monitoring, at $\gamma=5$, the no-jump weight decays more rapidly and the coherent oscillations are strongly suppressed, resulting in a more rapid, nearly monotonic drop in $\gamma_t$.

\bibliography{refsMIPT}

@article{Skinner19,
  title = {Measurement-Induced Phase Transitions in the Dynamics of Entanglement},
  author = {Skinner, Brian and Ruhman, Jonathan and Nahum, Adam},
  journal = {Phys. Rev. X},
  volume = {9},
  issue = {3},
  pages = {031009},
  numpages = {21},
  year = {2019},
  month = {Jul},
  publisher = {American Physical Society},
  doi = {10.1103/PhysRevX.9.031009},
  url = {https://link.aps.org/doi/10.1103/PhysRevX.9.031009}
}

@article{Elouard17,
	author = {Elouard, Cyril and Herrera-Mart{\'\i}, David A. and Clusel, Maxime and Auff{\`e}ves, Alexia},
	date = {2017/03/10},
	doi = {10.1038/s41534-017-0008-4},
	id = {Elouard2017},
	isbn = {2056-6387},
	journal = {npj Quantum Information},
	number = {1},
	pages = {9},
	title = {The role of quantum measurement in stochastic thermodynamics},
	url = {https://doi.org/10.1038/s41534-017-0008-4},
	volume = {3},
	year = {2017}}

@article{gemmer,
  author    = {J. Gemmer and M. Michel and G. Mahler 

},
  title     = {Quantum Thermodynamics},
  journal   = {Springer, Berlin},
  volume    = {},
  pages     = {},
  year      = {2004},
  doi       = {https://link.springer.com/book/10.1007/b98082},
  url       = {https://link.springer.com/book/10.1007/b98082},
}

@article{anders,
  author    = {Vinjanampathy, S. and Anders, J. 

},
  title     = {Quantum thermodynamics},
  journal   = {Contemporary Physics},
  volume    = {57(4)},
  pages     = {545–579},
  year      = {2016},
  doi       = {https://doi.org/10.1080/00107514.2016.1201896},
  url       = {https://www.tandfonline.com/doi/full/10.1080/00107514.2016.1201896},
}

@article{silva,
  author    = {A. Silva

},
  title     = {Statistics of the Work Done on a Quantum Critical System by Quenching a Control Parameter},
  journal   = {Phys. Rev. Lett},
  volume    = {101},
  pages     = {120603},
  year      = {2008},
  doi       = {https://doi.org/10.1103/PhysRevLett.101.120603
},
  url       = {https://journals.aps.org/prl/abstract/10.1103/PhysRevLett.101.120603},
}

@article{kosloff,
  author    = {R Kosloff 

},
  title     = {Quantum Thermodynamics: A Dynamical Viewpoint},
  journal   = {Entropy},
  volume    = {15(6)},
  pages     = {2100-2128},
  year      = {2013},
  doi       = {https://doi.org/10.3390/e15062100},
  url       = {https://www.mdpi.com/1099-4300/15/6/2100},
}

@article{goold,
  author    = {J Goold and M Huber and A Riera and L. del Rio and P. Skrzypczyk},
  title     = {The role of quantum information in thermodynamics—a topical review},
  journal   = {J. Phys. A: Math. Theor.},
  volume    = {49},
  pages     = {143001},
  year      = {2016},
  doi       = {10.1088/1751-8113/49/14/143001},
  url       = {https://iopscience.iop.org/article/10.1088/1751-8113/49/14/143001},
}

@article{mukamel,
  author    = {M Esposito and U Harbola and S Mukamel
},
  title     = {Nonequilibrium fluctuations, fluctuation theorems, and counting statistics in quantum systems},
  journal   = {Rev. Mod. Phys.},
  volume    = {86},
  pages     = {1125},
  year      = {2014},
  doi       = {https://doi.org/10.1103/RevModPhys.81.1665},
  url       = {https://journals.aps.org/rmp/abstract/10.1103/RevModPhys.81.1665},
}

@article{talkner,
  author    = {M Campisi and P Hänggi and P Talkner

},
  title     = {Colloquium: Quantum fluctuation relations: Foundations and applications},
  journal   = {Rev. Mod. Phys.},
  volume    = {83},
  pages     = {1653},
  year      = {2011},
  doi       = {https://doi.org/10.1103/RevModPhys.83.771},
  url       = {https://journals.aps.org/rmp/abstract/10.1103/RevModPhys.83.771},
}

@article{Seifert_2012,
doi = {10.1088/0034-4885/75/12/126001},
url = {https://dx.doi.org/10.1088/0034-4885/75/12/126001},
year = {2012},
month = {nov},
publisher = {IOP Publishing},
volume = {75},
number = {12},
pages = {126001},
author = {Seifert, Udo},
title = {Stochastic thermodynamics, fluctuation theorems and molecular machines},
journal = {Reports on Progress in Physics}
}

@article{jacobs,
  author    = {J. M. Horowitz and K. Jacobs
},
  title     = {Quantum effects improve the energy efficiency of feedback control},
  journal   = {Phys. Rev. E},
  volume    = {89},
  pages     = {042134},
  year      = {2014},
  doi       = {https://doi.org/10.1103/PhysRevE.89.042134},
  url       = {https://journals.aps.org/pre/abstract/10.1103/PhysRevE.89.042134},
}

@article{adesso,
  author    = {F Binder and L. A. Correa and C Gogolin and J. Anders and G Adesso },

  title     = {Thermodynamics in the Quantum Regime},
  journal   = {Springer, Cham,},
  volume    = {},
  pages     = {},
  year      = {2018},
  doi       = {https://link.springer.com/book/10.1007/978-3-319-99046-0},
  url       = {https://link.springer.com/book/10.1007/978-3-319-99046-0},
}

@article{kurchan,
  author    = {J Kurchan},

  title     = {A Quantum Fluctuation Theorem},
  journal   = {arXiv:cond-mat/0007360},
  volume    = {},
  pages     = {},
  year      = {2001},
  doi       = { 	
https://doi.org/10.48550/arXiv.cond-mat/0007360},
  url       = {https://arxiv.org/abs/cond-mat/0007360},
}

@article{milburn,
  author    = {H. M. Wiseman and G. J. Milburn},
  title     = {Quantum Measurement and Control},
  journal   = {Cambridge University Press},
  volume    = {},
  number    = {},
  pages     = {},
  year      = {2009},
  isbn      = {9780521804424},
  doi       = {10.1017/CBO9780511813948},
  url       = {https://doi.org/10.1017/CBO9780511813948},
}

@article{mishra,
  author    = {B. Misra and
E. C. G. Sudarshan},
  title     = {The Zeno’s paradox in quantum theory},
  journal   = {J. Math. Phys},
  volume    = {18},
  pages     = {756-763},
  year      = {1977},
  doi       = {
https://doi.org/10.1063/1.523304
},
  url       = {https://pubs.aip.org/aip/jmp/article/18/4/756/225634/The-Zeno-s-paradox-in-quantum-theory},
}

@article{Jarzynski97,
  title = {Nonequilibrium Equality for Free Energy Differences},
  author = {Jarzynski, C.},
  journal = {Phys. Rev. Lett.},
  volume = {78},
  issue = {14},
  pages = {2690--2693},
  numpages = {0},
  year = {1997},
  month = {Apr},
  publisher = {American Physical Society},
  doi = {10.1103/PhysRevLett.78.2690},
  url = {https://link.aps.org/doi/10.1103/PhysRevLett.78.2690}
}

@article{Lutz07,
  title = {Fluctuation theorems: Work is not an observable},
  author = {Talkner, Peter and Lutz, Eric and H\"anggi, Peter},
  journal = {Phys. Rev. E},
  volume = {75},
  issue = {5},
  pages = {050102},
  numpages = {2},
  year = {2007},
  month = {May},
  publisher = {American Physical Society},
  doi = {10.1103/PhysRevE.75.050102},
  url = {https://link.aps.org/doi/10.1103/PhysRevE.75.050102}
}

@article{Campisi11,
  title = {Colloquium: Quantum fluctuation relations: Foundations and applications},
  author = {Campisi, Michele and H\"anggi, Peter and Talkner, Peter},
  journal = {Rev. Mod. Phys.},
  volume = {83},
  issue = {3},
  pages = {771--791},
  numpages = {0},
  year = {2011},
  month = {Jul},
  publisher = {American Physical Society},
  doi = {10.1103/RevModPhys.83.771},
  url = {https://link.aps.org/doi/10.1103/RevModPhys.83.771}
}

@article{Gherardini22,
title = {Energy fluctuation relations and repeated quantum measurements},
journal = {Chaos, Solitons \& Fractals},
volume = {156},
pages = {111890},
year = {2022},
issn = {0960-0779},
doi = {https://doi.org/10.1016/j.chaos.2022.111890},
url = {https://www.sciencedirect.com/science/article/pii/S0960077922001011},
author = {Stefano Gherardini and Lorenzo Buffoni and Guido Giachetti and Andrea Trombettoni and Stefano Ruffo}
}

@article{Zhou21,
  title = {Work statistics in non-Hermitian evolutions with Hermitian endpoints},
  author = {Zhou, Zheng-Yang and Xiang, Ze-Liang and You, J. Q. and Nori, Franco},
  journal = {Phys. Rev. E},
  volume = {104},
  issue = {3},
  pages = {034107},
  numpages = {11},
  year = {2021},
  month = {Sep},
  publisher = {American Physical Society},
  doi = {10.1103/PhysRevE.104.034107},
  url = {https://link.aps.org/doi/10.1103/PhysRevE.104.034107}
}

@article{Deffner15,
  title = {Jarzynski Equality in PT-Symmetric Quantum Mechanics},
  author = {Deffner, Sebastian and Saxena, Avadh},
  journal = {Phys. Rev. Lett.},
  volume = {114},
  issue = {15},
  pages = {150601},
  numpages = {5},
  year = {2015},
  month = {Apr},
  publisher = {American Physical Society},
  doi = {10.1103/PhysRevLett.114.150601},
  url = {https://link.aps.org/doi/10.1103/PhysRevLett.114.150601}
}

@article{Wei18,
  title = {Quantum work relations and response theory in parity-time-symmetric quantum systems},
  author = {Wei, Bo-Bo},
  journal = {Phys. Rev. E},
  volume = {97},
  issue = {1},
  pages = {012114},
  numpages = {7},
  year = {2018},
  month = {Jan},
  publisher = {American Physical Society},
  doi = {10.1103/PhysRevE.97.012114},
  url = {https://link.aps.org/doi/10.1103/PhysRevE.97.012114}
}

@article{Deffner16,
	author = {Gardas, Bart{\l}omiej and Deffner, Sebastian and Saxena, Avadh},
	date = {2016/03/22},
	doi = {10.1038/srep23408},
	id = {Gardas2016},
	isbn = {2045-2322},
	journal = {Scientific Reports},
	number = {1},
	pages = {23408},
	title = {Non-hermitian quantum thermodynamics},
	url = {https://doi.org/10.1038/srep23408},
	volume = {6},
	year = {2016}}

@article{Zeng_2017,
doi = {10.1088/2399-6528/aa8f26},
url = {https://dx.doi.org/10.1088/2399-6528/aa8f26},
year = {2017},
month = {nov},
publisher = {IOP Publishing},
volume = {1},
number = {3},
pages = {031001},
author = {Zeng, Meng and Yong, Ee Hou},
title = {Crooks fluctuation theorem in PT-symmetric quantum mechanics},
journal = {Journal of Physics Communications}
}

@article{PrasannaVenkatesh_2014,
doi = {10.1088/1367-2630/16/1/015032},
url = {https://dx.doi.org/10.1088/1367-2630/16/1/015032},
year = {2014},
month = {jan},
publisher = {IOP Publishing},
volume = {16},
number = {1},
pages = {015032},
author = {Prasanna Venkatesh, B and Watanabe, Gentaro and Talkner, Peter},
title = {Transient quantum fluctuation theorems and generalized measurements},
journal = {New Journal of Physics}
}

@article{Watanabe2,
  title = {Generalized energy measurements and modified transient quantum fluctuation theorems},
  author = {Watanabe, Gentaro and Venkatesh, B. Prasanna and Talkner, Peter},
  journal = {Phys. Rev. E},
  volume = {89},
  issue = {5},
  pages = {052116},
  numpages = {7},
  year = {2014},
  month = {May},
  publisher = {American Physical Society},
  doi = {10.1103/PhysRevE.89.052116},
  url = {https://link.aps.org/doi/10.1103/PhysRevE.89.052116}
}

@article{Watanabe14,
  title = {Quantum fluctuation theorems and generalized measurements during the force protocol},
  author = {Watanabe, Gentaro and Venkatesh, B. Prasanna and Talkner, Peter and Campisi, Michele and H\"anggi, Peter},
  journal = {Phys. Rev. E},
  volume = {89},
  issue = {3},
  pages = {032114},
  numpages = {8},
  year = {2014},
  month = {Mar},
  publisher = {American Physical Society},
  doi = {10.1103/PhysRevE.89.032114},
  url = {https://link.aps.org/doi/10.1103/PhysRevE.89.032114}
}

@article{Rastegin_2013,
doi = {10.1088/1742-5468/2013/06/P06016},
url = {https://dx.doi.org/10.1088/1742-5468/2013/06/P06016},
year = {2013},
month = {jun},
publisher = {IOP Publishing and SISSA},
volume = {2013},
number = {06},
pages = {P06016},
author = {Rastegin, Alexey E},
title = {Non-equilibrium equalities with unital quantum channels},
journal = {Journal of Statistical Mechanics: Theory and Experiment}
}

@article{Albash13,
  title = {Fluctuation theorems for quantum processes},
  author = {Albash, Tameem and Lidar, Daniel A. and Marvian, Milad and Zanardi, Paolo},
  journal = {Phys. Rev. E},
  volume = {88},
  issue = {3},
  pages = {032146},
  numpages = {14},
  year = {2013},
  month = {Sep},
  publisher = {American Physical Society},
  doi = {10.1103/PhysRevE.88.032146},
  url = {https://link.aps.org/doi/10.1103/PhysRevE.88.032146}
}

@article{Rastegin14,
  title = {Jarzynski equality for quantum stochastic maps},
  author = {Rastegin, Alexey E. and \ifmmode \dot{Z}\else \.{Z}\fi{}yczkowski, Karol},
  journal = {Phys. Rev. E},
  volume = {89},
  issue = {1},
  pages = {012127},
  numpages = {10},
  year = {2014},
  month = {Jan},
  publisher = {American Physical Society},
  doi = {10.1103/PhysRevE.89.012127},
  url = {https://link.aps.org/doi/10.1103/PhysRevE.89.012127}
}

@article{Aberg18,
  title = {Fully Quantum Fluctuation Theorems},
  author = {\AA{}berg, Johan},
  journal = {Phys. Rev. X},
  volume = {8},
  issue = {1},
  pages = {011019},
  numpages = {78},
  year = {2018},
  month = {Feb},
  publisher = {American Physical Society},
  doi = {10.1103/PhysRevX.8.011019},
  url = {https://link.aps.org/doi/10.1103/PhysRevX.8.011019}
}

@article{turkeshi_4,
  author    = {X. Turkeshi and M. Dalmonte and R. Fazio and M. Schirò},
  title     = {Entanglement transitions from stochastic resetting of non-Hermitian quasiparticles},
  journal   = {Phys. Rev. B},
  volume    = {107},
  pages     = {079901},
  year      = {2023},
  doi       = {10.1103/PhysRevB.105.L241114},
  url       = {https://doi.org/10.1103/PhysRevB.105.L241114},
}

@article{turkeshi_5,
  author    = {X. Turkeshi and A. Biella and R. Fazio and M. Dalmonte and M. Schiró},
  title     = {Measurement-induced entanglement transitions in the quantum Ising chain: From infinite to zero clicks},
  journal   = {Phys. Rev. B},
  volume    = {103},
  pages     = {224210},
  year      = {2021},
  doi       = {10.1103/PhysRevB.103.224210},
  url       = {https://doi.org/10.1103/PhysRevB.103.224210},
}

@article{ueda0,
  author    = {Y. Ashida and Z. Gong and M. Ueda},
  title     = {Non-Hermitian physics},
  journal   = {Advances in Physics},
  volume    = {69},
  number    = {3},
  pages     = {249--435},
  year      = {2020},
  doi       = {10.1080/00018732.2021.1876991},
  url       = {https://doi.org/10.1080/00018732.2021.1876991},
}

@article{silva_2,
  author    = {C. Zerba and A. Silva},
  title     = {Measurement phase transitions in the no-click limit as quantum phase transitions of a non-hermitean vacuum},
  journal   = {SciPost Phys. Core},
  volume    = {6},
  pages     = {051},
  year      = {2023},
  doi       = {10.21468/SciPostPhysCore.6.3.051},
  url       = {https://www.scipost.org/SciPostPhysCore.6.3.051?acad_field_slug=politicalscience},
}

@article{wiseman,
  author    = {H. M. Wiseman},
  title     = {Quantum trajectories and quantum measurement theory},
  journal   = {Quantum Semiclass. Opt.},
  volume    = {8},
  pages     = {205},
  year      = {1996},
  url       = {https://iopscience.iop.org/article/10.1088/1355-5111/8/1/015},
}

@article{ueda,
  author    = {R. Hamazaki and K. Kawabata and N. Kura and M. Ueda},
  title     = {Universality classes of non-Hermitian random matrices},
  journal   = {Phys. Rev. Research},
  volume    = {2},
  pages     = {023286},
  year      = {2020},
  doi       = {10.1103/PhysRevResearch.2.023286},
  url       = {https://doi.org/10.1103/PhysRevResearch.2.023286},
}

@article{Bruzewicz2019,
  author    = {C. D. Bruzewicz and J. Chiaverini and R. McConnell and J. M. Sage},
  title     = {Trapped-ion quantum computing: Progress and challenges},
  journal   = {Appl. Phys. Rev.},
  volume    = {6},
  pages     = {021314},
  year      = {2019},
  doi       = {10.1063/1.5088164},
  url       = {https://doi.org/10.1063/1.5088164},
}

@article{Gaebler2022,
  author    = {J. P. Gaebler and others},
  title     = {Randomized benchmarking of multiqubit gates with arbitrary controlled-phase operations in a trapped-ion quantum processor},
  journal   = {Phys. Rev. Lett.},
  volume    = {129},
  pages     = {240501},
  year      = {2022},
  doi       = {10.1103/PhysRevLett.129.240501},
  url       = {https://doi.org/10.1103/PhysRevLett.129.240501},
}

@article{Bluvstein2022,
  author    = {A. Bluvstein and others},
  title     = {A logical qubit in a neutral atom array with programmable {Rydberg} interactions},
  journal   = {Nature},
  volume    = {604},
  pages     = {451},
  year      = {2022},
  doi       = {10.1038/s41586-022-04592-6},
  url       = {https://doi.org/10.1038/s41586-022-04592-6},
}

@article{Kim2023,
  author    = {Y. Kim and others},
  title     = {Evidence for the utility of quantum computing before fault tolerance},
  journal   = {Nature},
  volume    = {618},
  pages     = {500},
  year      = {2023},
  doi       = {10.1038/s41586-023-06096-3},
  url       = {https://doi.org/10.1038/s41586-023-06096-3},
}

@article{Haase2024,
  author    = {J. F. Haase and P. A. Ivanov and F. Schmidt-Kaler and T. Schätz},
  title     = {Quantum sensing with entangled atoms: From foundational studies to practical applications},
  journal   = {Rev. Mod. Phys.},
  volume    = {96},
  pages     = {015002},
  year      = {2024},
  doi       = {10.1103/RevModPhys.96.015002},
  url       = {https://doi.org/10.1103/RevModPhys.96.015002},
}

@article{Endo2021,
  author    = {S. Endo and Z. Cai and S. C. Benjamin and X. Yuan},
  title     = {Hybrid quantum-classical algorithms and quantum error mitigation},
  journal   = {J. Phys. Soc. Jpn.},
  volume    = {90},
  pages     = {074001},
  year      = {2021},
  doi       = {10.7566/JPSJ.90.074001},
  url       = {https://doi.org/10.7566/JPSJ.90.074001},
}

@article{Dorner2013,
  author    = {R. Dorner and S. R. Clark and L. Heaney and R. Fazio and J. Goold and V. Vedral},
  title     = {Extracting quantum work statistics and fluctuation theorems by single-qubit interferometry},
  journal   = {Phys. Rev. Lett.},
  volume    = {110},
  pages     = {230601},
  year      = {2013},
  doi       = {10.1103/PhysRevLett.110.230601},
  url       = {https://doi.org/10.1103/PhysRevLett.110.230601},
}

@article{Mazzola2013,
  author    = {L. Mazzola and G. De Chiara and M. Paternostro},
  title     = {Measuring the Characteristic Function of the Work Distribution},
  journal   = {Phys. Rev. Lett.},
  volume    = {110},
  pages     = {230602},
  year      = {2013},
  doi       = {10.1103/PhysRevLett.110.230602},
  url       = {https://doi.org/10.1103/PhysRevLett.110.230602},
}

@article{Batalhao2014,
  author    = {T. B. Batalh{\~a}o and A. M. Souza and L. Mazzola and R. Auccaise and R. S. Sarthour and I. S. Oliveira and J. Goold and G. De Chiara and M. Paternostro and R. M. Serra},
  title     = {Experimental Reconstruction of Work Distribution and Study of Fluctuation Relations in a Closed Quantum System},
  journal   = {Phys. Rev. Lett.},
  volume    = {113},
  pages     = {140601},
  year      = {2014},
  doi       = {10.1103/PhysRevLett.113.140601},
  url       = {https://doi.org/10.1103/PhysRevLett.113.140601},
}

@article{An2015,
  author    = {S. An and J.-N. Zhang and M. Um and D. Lv and Y. Lu and J. Zhang and Z.-Q. Yin and H. T. Quan and K. Kim},
  title     = {Experimental test of the quantum Jarzynski equality with a trapped-ion system},
  journal   = {Nat. Phys.},
  volume    = {11},
  pages     = {193--199},
  year      = {2015},
  doi       = {10.1038/nphys3197},
  url       = {https://doi.org/10.1038/nphys3197},
}

@article{tasaki2000,
  author    = {Hal Tasaki},
  title     = {Jarzynski Relations for Quantum Systems and Some Applications},
  journal   = {arXiv preprint cond-mat/0009244},
  year      = {2000},
  doi       = {10.48550/arXiv.cond-mat/0009244},
  url       = {https://doi.org/10.48550/arXiv.cond-mat/0009244},
}

\end{document}